\documentclass[<1p>]{elsarticle} 

\usepackage{amsmath, amsthm, amssymb, amsthm}
\usepackage{subcaption}
\usepackage[flushleft]{threeparttable}
\usepackage{tabu,multirow}
\usepackage{color}
\usepackage{etoolbox}
\usepackage{float}
\usepackage{booktabs}
\usepackage{etoolbox}
\let\today\relax
\usepackage{etoolbox}
\usepackage{etoolbox}
\usepackage{booktabs}
\usepackage{siunitx}
\patchcmd{\MaketitleBox}{\footnotesize\itshape\elsaddress\par\vskip36pt}{\footnotesize\itshape\elsaddress\par\parbox[b][36pt]{\linewidth}{\vfill\hfill\textnormal{\today}\hfill\null\vfill}}{}{}%
\patchcmd{\pprintMaketitle}{\footnotesize\itshape\elsaddress\par\vskip36pt}{\footnotesize\itshape\elsaddress\par\parbox[b][36pt]{\linewidth}{\vfill\hfill\textnormal{\today}\hfill\null\vfill}}{}{}%
\makeatletter
\def\ps@pprintTitle{%
    \let\@oddhead\@empty
    \let\@evenhead\@empty
    \def\@oddfoot{\footnotesize\itshape
    {November 2022}\hfill\today}%
    \let\@evenfoot\@oddfoot
    }
\makeatother
\begin{document}
\begin{frontmatter}
\begin{abstract}
This paper proposes strategies to detect time reversibility in stationary stochastic processes by 
using the properties of mixed causal and noncausal models. It shows that they can also be used 
for non-stationary processes when the trend component is computed with the Hodrick-Prescott filter 
rendering a time-reversible closed-form solution. This paper also links the concept of an environmental 
tipping point to the statistical property of time irreversibility and assesses fourteen climate indicators. 
We find evidence of time irreversibility in $GHG$ emissions, global temperature, global sea levels, sea ice 
area, and some natural oscillation indices. While not conclusive, our findings urge the implementation 
of correction policies to avoid the worst consequences of climate change and not miss the opportunity 
window, which might still be available, despite closing quickly.
\end{abstract}
\title{Is climate change time-reversible?}
\date{\today}
\author[a]{Francesco Giancaterini \footnote{Corresponding author:
Francesco Giancaterini, Maastricht University, School of Business and Economics, Department
of Quantitative Economics, P.O.box 616, 6200 MD, Maastricht, The Netherlands.\\Email: f.giancaterini@maastrichtuniversity.nl.}}
\author[a]{Alain Hecq}
\author[b]{Claudio Morana}
\address[a]{Maastricht University}
\address[b]{University of Milano-Bicocca\\
Center for European Studies -Milan\\
RCEA, RCEA-Europe ETS\\
CeRP - Collegio Carlo Alberto}
\graphicspath{ {Pictures/} }

\end{frontmatter}

\section{Introduction}
According to the most recent International Panel on Climate Change report, humanity is unlikely to prevent global
warming by $1.5^{\circ}$ above pre-industrial levels. Still, aggressive curbing of greenhouse-gas emissions and 
carbon extraction from the atmosphere could limit its rise and even bring it back down (\cite{IPCC22}). But this 
window is rapidly closing, and, above the $1.5^{\circ}$ threshold, the chances of tipping points, extreme weather, 
and ecosystem collapse will become even more sizeable.\\
\indent An environmental tipping point is when small climatic changes might trigger large, abrupt, and irreversible 
environmental changes and lead to cascading effects. Recent IPCC assessments suggest that tipping points 
might arise between $1^{\circ}$ and $2^{\circ}$ warming, and likely to manifest at current emissions levels if 
they have not already occurred. Well-known tipping points concern the Greenland and the West Antarctic ice 
sheets, the Atlantic Meridional Overturning Circulation ($AMOC$), thawing permafrost, $ENSO$, and the Amazon 
rainforest. Recent evidence suggests that melting ice sheets is accelerating because of warming air and ocean 
temperatures and less snowfall. Some studies indicate that the irreversible disintegration of the Greenland ice sheet 
could occur at $0.8^{\circ}$ and $3.2^{\circ}$ warming (\cite{wunderling2021interacting}). An unstoppable ice 
sheet melting in Antarctica would manifest at $2^{\circ}$ warming (\cite{deconto2021paris}). Ice sheets melting adds 
fresh water to the North Atlantic, weakening the $AMOC$, one of the main global ocean currents, which is 
already in its weakest state in 1,000 years (\cite{caesar2021current}). Its shutdown would cause significant 
cooling along the US east coast and Western Europe, alter rainfall and cause more drying. 
At the current global warming pace, a $50\%$ weakening of $AMOC$ is expected by 2100, and a tipping point
between $3^{\circ}$ and $5.5^{\circ}$ warming. Moreover, the Arctic is warming twice as faster as the planet on 
average, and it has already warmed $2^{\circ}$, causing permafrost thawing, which releases CO2 and methane 
into the atmosphere. Available estimates point to 1400 billion tons of carbon frozen in the Arctic’s permafrost, 
twice as much carbon already in the atmosphere, and a $2^{\circ}$ warming could even cause the thawing of 
40$\%$ of the world’s permafrost. The El Ni{\~n}o-Southern Oscillation or $ENSO$ cycle is an oscillating warming 
and cooling pattern affecting rainfall intensity and temperatures in tropical regions. It can strongly influence 
weather in many parts of the globe. El Ni{\~n}o and La Ni{\~n}a are the warm and cool phases of the $ENSO$ 
cycle, respectively. Oceans warming can trigger a tipping point in the $ENSO$ cycle, increasing its variability and 
intensity and shifting its teleconnection eastward (\cite{cai2021changing}). Extreme rainfalls and droughts will no 
longer occur in tropical regions but throughout the earth due to the destabilization of these natural oscillations.
The Amazon rainforest has already lost about $17\%$ of its tree cover. At the current rate of deforestation, 
the loss could reach $27\%$ by 2030. \cite{lovejoy2018amazon} estimate the dieback of the Amazon Forest at 
$20\%$-$25\%$; beyond this deforestation threshold, the rainforest would transform into a savannah, potentially 
releasing up to 90 gigatons of CO2. Some climate models already indicate that the Amazon will be a net generator 
of C02 by 2035, setting the dieback threshold at $3^{\circ}$ warming.\\
\indent Further uncertainty on the compound effect of the above phenomena arises from their potential interaction,
allowing tipping points to occur even below $2^{\circ}$ warming. Overall, greenhouse gases generated by human
 activity over the last two centuries have driven the global trend temperature up. This temperature warming has 
 widely impacted the natural environment and has raised the risk of irreversible changes of state with catastrophic 
 consequences (see also \cite{schellnhuber2008global}) and \cite{solomon2009irreversible}).\\
\indent In this paper, we link the concept of an environmental tipping point to the statistical property of time 
irreversibility. A stationary process $\{ Y_{t} \}_{t=1}^{T} $ is said to be time-reversible if its statistical properties 
are independent of the direction of time. In other words, the vectors $(Y_{1}, Y_{2}, \dots, Y_{T})$ have identical 
joint distributions as $(Y_{-T}, Y_{-(T-1)}, \dots , Y_{-1})$ for every integer $T$. Hence, a time-reversible process 
(TR) exhibits a temporal symmetry in its probabilistic structure. In the alternative circumstance, we have time 
irreversibility when the stochastic process behaves differently according to the direction of time considered. TR 
has been under investigation in various fields over the years, for instance, in the different branches of physics, 
where researchers have been investigating whether time has some preferred direction in explaining physical 
phenomena (see \cite{wald1980quantum}, \cite{levesque1993molecular}, \cite{holster2003criterion}). This 
univocity along the time direction appears to be a tipping point property, as once a tipping point is reached, the 
system undergoes an irreversible state change.\\
\indent This paper aims to investigate whether TR has the potential to offer insight into the process of climate 
change and its implications for the natural environment. Studying TR in the context 
of climate change is motivated by the possibility of answering the following questions: are there divergences 
between the forward-time and backward-time joint probability distributions for the process of climate change and 
global warming? Are these processes symmetric over time? Is this property similarly present in natural oscillations
that temperature warming might have permanently impacted, inducing changes in their frequencies and intensity
of occurrence? Irreversibility in this context might carry insights into the event of state changes.\\
\indent This paper then introduces new strategies to detect whether a stochastic process is time-reversible. 
There are already several tests for TR in the econometric literature. See, for instance, \cite{ramsey1996time}, 
\cite{hinich1998frequency}, \cite{chen2000testing}, \cite{belaire2003tests}, and \cite{proietti2020peaks}. The 
shortcoming of many of these approaches is that they usually impose strong restrictions on the model or are not 
trivial to apply. Our new strategies are grounded on mixed causal and noncausal models (see 
\cite{gourieroux2016filtering}). Unlike causal models, which only consider the relationship between present and 
lagged values, mixed casual and noncausal models also compute the relationship between present and future 
values. This framework leads to nonlinear conditional expectations (e.g., \cite{gourieroux2022nonlinear}). The 
connection between these models and TR gives rise to our testing strategies.\\
\indent Furthermore, similarly to \cite{proietti2020peaks}, we can test for TR on non-stationary time series using 
a novel approach. We extract the trend component using the Hodrick-Prescott (HP) filter imparted in a 
time-reversible closed-form solution. Then, the cyclical component, which records the process's oscillations 
around its trend, is responsible for the potential time-irreversibility feature of the stochastic process.\\
\indent The rest of the paper is as follows. Section 2 summarizes the properties of time-reversible processes 
and reviews the existing methods to detect TR. Section 3 introduces our new TR strategies. Namely, we show 
how our new approaches exploit the properties of mixed causal and noncausal models. We then evaluate their 
performance through Monte Carlo experiments. Section 4 extends our framework to non-stationary time series, 
and Section 5 presents the empirical assessment of some relevant climate variables. Finally, Section 6 concludes.

\section{Time reversibility}
\cite{weiss1975time} shows that if a Gaussian error term characterizes an ARMA model, then the process is 
time-reversible. Indeed, Gaussian processes are entirely defined by their second-order moments, which 
have the property of being time symmetrical.\\
\indent \cite{hallin1988time} consider two-sided linear models of the form:
\begin{equation}
Y_{t}= \sum_{k=- \infty}^{\infty} \theta_{k}\epsilon_{t-k},
\end{equation}
where the stationary condition $\sum_{k=- \infty}^{\infty} |\theta_{k}| < \infty$ is satisfied. They claim that if 
$ \{ Y_{t} \}_{t=1}^{T} $ is time-reversible, then either $ \epsilon_{t}$ is a Gaussian white noise, or there exists 
a $k$ and $s \in \{ 0, 1 \}$ such that $\theta_{2k + j}=(-1)^{s} \theta_{2k - j}$. However, $\epsilon_{t}$ has to be 
a sequence of $i.i.d.$ zero-mean random variables with finite moments of all orders. It is an unrealistic 
assumption for non-Gaussian processes and many time series.\\
\indent \cite{breidt1992time} extend Weiss's results to non-Gaussian processes assuming milder conditions 
than  \cite{hallin1988time}. They take the following $ARMA(p,q)$ process into account:
\begin{equation}
\phi (L) Y_{t}= \theta (L)\epsilon_{t},
\end{equation}
where $L$ indicates the backshift operator, $\phi(z)$ has $r$ roots outside and $s$ roots inside 
the unit circle ($r+s=p$), and $\epsilon_{t}$ has a finite variance. For simplicity, we set the polynomial 
$\theta (L)=1$, such that (2) can be rewritten as:
\begin{equation}
\phi^{+}(L) \phi^{-}(L)Y_{t}=\epsilon_{t},
\end{equation}
where $\phi^{+}(L)$ has $r$ roots outside the unit circle while $\phi^{-}(L)$ has $s$ roots inside. It is well 
known that (3) has a unique stationary solution given by a two-sided moving average representation, as 
expressed in (1). \cite{breidt1992time} claim that if $\phi (z)$ and $\phi (z^{-1})$ have different roots, then 
${ Y_{t} }$ is reversible if and only if the error term is Gaussian. In the other case, that is when the two 
polynomials $\phi (z)$ and $\phi (z^{-1})$ have the same roots, (1) (or equivalently (3)) is time-reversible 
regardless of the distribution of $\epsilon_{t}$. Indeed, if $p >0$ and $\phi(z)$ and $\phi (z^{-1})$ have the 
same roots, $1/\phi(z)$ has the Laurent expansion
\begin{equation}
\frac{1}{\phi (z)}=\sum_{- \infty}^{\infty} \theta_{j} z^{j},
\end{equation}
with $\theta_{-p/2 - j}=\theta_{-p/2 + j}$, for $j=0, 1, \dots $ (see \cite{breidt1992time}). 
This implies that the result of \cite{hallin1988time} is a consequence of the conclusion that the two polynomials 
$\phi(z)$ and $\phi(z^{-1})$ have the same roots. Moreover, unlike \cite{hallin1988time}, \cite{breidt1992time},
only assume that the error term must have finite variance.\\
\indent \cite{ramsey1996time} define the stationary stochastic process $ \{ Y_{t} \}_{t=1}^{T} $ is time-reversible 
only if:
\begin{equation} 
\gamma_{i,j}=E[Y_{t}^{i}Y_{t-k}^{j}]-E[Y_{t}^{j}Y_{t-k}^{i}]=0
\end{equation}
for all $i, j, k$ $\in$ $\mathbb{N}^{+}$. This is a sufficient condition for TR, but not a necessary one since it only 
considers a proper subset of moments from the joint distributions of $\{Y_{t}\}$. Since it is impractical to show 
that (5) holds for any $i$, $j$, and $k$, they adopt a restricted definition of TR by imposing $i+k \leq m$ and 
$k \leq K$. In particular, they restrict $m=3$ so that the symmetric-bicovariance function is given by:
\begin{equation}
\gamma_{2,1}=E[Y_{t}^{2}Y_{t-k}]-E[Y_{t}Y_{t-k}^{2}]=0,
\end{equation}
for all integer values of $k$. \cite{ramsey1996time} claim that $i+j=3$ is sufficient to provide a valid indication of 
time 
irreversibility.\\
\indent \cite{ramsey1996time} also introduced a new procedure to test TR that became a standard approach 
to investigating business cycle properties such as asymmetry. It amounts to a TR test statistic distributed as a 
standard normal distribution:
\begin{equation}
\sqrt{T}\frac{[\widehat{\gamma}_{2,1} - \gamma_{2,1}]}{\sqrt{Var(\widehat{\gamma}_{2,1})}} \xrightarrow{d} N(0,1),
\end{equation}
with:
\begin{equation*}
\widehat{\gamma}_{2,1}=\widehat{B}_{2,1}(k) - \widehat{B}_{1,2}(k),
\end{equation*}
and:
\begin{equation*}
\widehat{B}_{2,1}=(T-k)^{-1} \sum_{t=K+1}^{T}Y_{t}^{2} Y_{t-k} \ ; \ \ \ \widehat{B}_{1,2}=(T-k)^{-1} \sum_{t=K+1}^{T}Y_{t} Y_{t-k}^{2},
\end{equation*}
for various integer values of $k$. Under the null hypothesis, we have a time-reversible process. The pre-requisite 
of the test is that the data must possess finite first sixth moment. If the distribution lacks this property, the test size
can be seriously distorted (see \cite{belaire2003tests}).\\
\indent \cite{chen2000testing} propose a new class of TR tests, which, unlike \cite{ramsey1996time}, does not 
require any moment restrictions. This class of tests relies on the fact that if $\{ Y_{t} \}_{t=1}^{T}$ is a 
time-reversible process, then for every $k=1, 2, \dots$, the distribution of $X_{t,k}=Y_{t}-Y_{t-k}$ is symmetric 
about the origin. The drawback of this approach is that it allows for testing the symmetry of $X_{t,k}$ for each 
value of $k$, but not jointly for a collection of $k$ values, which would require a portmanteau 
test.\footnote{ \cite{chen2000testing} state that to jointly test $X_{t,k}$ 
for a collection of $k$ values, a portmanteau test is required.} Moreover, its implementation is not trivial.\\
\indent Similar reasoning is followed by \cite{proietti2020peaks} since also his test is based on the idea that 
$X_{t,k}$ has to be symmetric for every $k>0$. However, \cite{proietti2020peaks} uses a weaker definition of 
TR as $ \{ Y_{t} \}_{t=1}^{T} $ can also be non-stationary.

\section{New strategies to detect time reversibility on stationary time series}
This Section introduces new strategies to assess TR in stationary stochastic processes, exploiting the properties 
of mixed causal and noncausal models. \cite{breid1991maximum} introduce mixed causal and noncausal models 
as expressed in equation (3). They define the polynomial $\phi^{-}(z)$ as noncausal and the polynomial 
$\phi^{+}(z)$ as causal. A required condition for identifying the causal from the noncausal component is the 
non-Gaussianity of the innovation term.\\
\indent \cite{lanne2011noncausal}, rewriting the noncausal polynomial in (3) as a lead polynomial, start with a 
mixed causal and noncausal model expressed as:
\begin{equation}
\phi (L) \varphi (L^{-1})Y_{t}=\epsilon_{t},
\end{equation}
where $L^{-1}$ produces lead such that $L^{-1}Y_{t}=Y_{t+1}$. A mixed causal and noncausal model 
represented in this way is denoted as MAR($r$,$s$), where $\varphi(L^{-1})$ is the noncausal polynomial of 
order $s$ and $\phi(L)$ is the causal polynomial of order $r$. Exactly as representation (3), $r+s=p$ is true 
even in this case. Purely causal and purely noncausal models are obtained setting respectively 
$\varphi(L^{-1})=1$ and $\phi(L)=1$ (see \cite{gourieroux2013explosive}, \cite{hencic2015noncausal}, 
\cite{hecq2016identification}, \cite{fries2019mixed}, \cite{hecq21predicting}, \cite{GIANCATERINI2022}, 
and \cite{fries2021conditional}). In (8), both causal and noncausal polynomials have their roots outside the 
unit circle, such that:
\begin{equation}
\phi(z) \neq 0 \ \ and \ \ \varphi(z) \neq 0 \ \ for \ \ |z| \leq 1.
\end{equation} 
\indent The tests for TR that we propose have the common feature of extending the results obtained by \cite{breidt1992time} to 
the MAR($r$,$s$) representation (8). This is possible if and only if $Condition \ 3.1$ is true.\\
\newline
\textbf{Condition 3.1} A stochastic process that can be expressed as a MAR model is time-reversible if and only if $\phi(z)\varphi(z^{-1})$ 
have the same roots as $\phi(z^{-1})\varphi(z)$.
Namely, when:
\begin{equation*}
r=s \ and \ \phi_{i}=\varphi_{i}, \ for \ \ i=1,\dots,s. \\
\end{equation*}
\newline
\indent This implies that MARs are time-reversible if and only if the causal polynomial has the same order and 
the same coefficients as the noncausal polynomial and vice versa. Remember that it is impossible to identify a 
MAR model under the Gaussianity of the innovation term. Hence, in that case, we have a time-reversible process 
(see \cite{weiss1975time}). 

\subsection{Strategy 1: For detecting time reversibility}
The first strategy aims to evaluate whether a stochastic process meets $Condition \ 3.1$. In particular, it 
uses a procedure similar to the one used to identify MAR models (see \cite{lanne2011noncausal} and 
\cite{hecq2016identification}). The procedure is as follows:
\begin{enumerate}
	\item We estimate a conventional autoregressive process (also called pseudo-causal model) by OLS, and 
	the lag order $p$ is selected using information criteria (for instance, AIC or BIC).
	
	\item We test the normality in the residuals of the AR($p$). If the null hypothesis of Gaussianity is not 
	rejected, we cannot identify a MAR($r$,$s$) model, and for the reasons above, we have a time-reversible 
	process. Moreover, if the null hypothesis of normality is rejected and the estimated $p$ is an odd number, 
	the condition $r = s$ can never be satisfied. According to $Condition \ 3.1$, this result would allow us to 
	identify our process as time-irreversible. However, the selection of $p$ might not be univocal and depend 
	on the information criterion employed. As such, to have more robust results before proceeding to the next 
	step, we increase $p$ by one unit so that $r=s$ is still possible. In the alternative case that $p$ is an even 
	number, we directly proceed to the next step.
		
	\item We select a model among all MAR($r$,$s$) specifications with $r+s=p$ if $p$ is an even number;
	otherwise $r+s=p+1$. This step is performed using a maximum likelihood approach (see
	\cite{GIANCATERINI2022} and references therein). In the selection procedure, we also include the model 
	given by the restricted likelihood that imposes commonalities in causal and noncausal parameters (the 
	model with the same restrictions as in $Condition \ 3.1$). Note that when we compute the information 
	criteria of the model with restricted likelihood, instead of estimating $p$ parameters (or $p+1$ if $p$ is an 
	odd number), we estimate $p/2$ of them (or $(p+1)/2$), implying a smaller penalty term. Finally, we 
	choose the model with the smallest information criteria.\\
\end{enumerate}
Consider a short example to illustrate how the strategy works. We suppose that we estimate a conventional 
AR model by OLS, and we reject the Gaussian hypothesis of the residuals, for instance, using the Jarque-Bera 
test. Furthermore, we assume we select the number of lags $p$ equal to 2. To analyze whether our process is 
time-reversible, we then compute the log-likelihoods and then the information criteria of the following four models: 
MAR(2,0), MAR(0,2), MAR(1,1) as well as the MAR(1,1) with the restriction $\phi=\varphi$. If the model with the 
smallest information criteria is the one with the restriction, we have a time-reversible process. 
We have a time-irreversible process in the alternative case where another model is selected. This approach 
allows knowing with a limited number of steps whether the process is time-reversible. Its shortcoming is that 
information criteria are very sensitive to the sample size, and model selection might not be robust to sample update or trimming. Moreover, model selection can depend on the information criterion employed, i.e., AIC rather than BIC, HQ, or others. Finally, even for the same information criterion, the value used for model selection can only slightly differ from values shown by either lower or higher-order alternative models.

\subsection{Strategy 2: For detecting time reversibility}
The second strategy we introduce is more robust concerning the sample and slight differences in the value of 
information criteria when models are compared. However, more steps are required to identify the TR of the 
process than for the previous approach. It requires the following steps: steps 1 and 2 are identical to what we 
described in 3.1;
\begin{enumerate}
	\setcounter{enumi}{2}
	\item We select a model among all MAR($r$,$s$) specifications with $r+s=p$ if $p$ is an even number 
	(otherwise $r+s=p+1$). Then, we choose the one with the largest likelihood (since we are considering 
	models with the same number of parameters). 
	
	\item If the selected model is the one with $r=s$ (in our previous example, it was the MAR(1,1)), we 
	compute a likelihood ratio test, taking into account the same restrictions as in $Condition \ 3.1$. 
	If we do not reject the test's null hypothesis, we have TR. On the other hand, if we reject the null 
	hypothesis, we identify the process under investigation as time-irreversible.
\end{enumerate}

\subsection{Simulation study}
We now analyze the performance of these two strategies using Monte Carlo experiments. We take into 
account data-generating processes ($dgp$) defined by an error term with a skewed Student's-$t$ distribution, 
generated by joining two scaled halves of the Student's-t distribution (see \cite{fernandez1998bayesian}): 
\begin{equation}
	f(\epsilon)= \frac{2}{\gamma + \frac{1}{\gamma}} \Biggl\{ g\Biggl(\frac{\epsilon}{\gamma} \Biggr)
	\mathcal{I}(\epsilon) + g\bigl(\gamma \epsilon\bigr) \mathcal{I}(- \epsilon) \Biggr\},
\end{equation} 
where $\mathcal{I} (\epsilon)$ and $\mathcal{I} (-\epsilon)$ stand for the indicator function:
\begin{center}
$\mathcal{I} (\epsilon)=
\begin{cases}
1, \ \ \ \epsilon \geq 0 \\
0, \ \ \ \epsilon < 0\\
\end{cases},$
\end{center}
$g(\epsilon)$ stands for the density function of a symmetric Student's-$t$, and $\gamma \in \mathbb{R}^{+}$. 
In case $\gamma = 1$, we have $f(\epsilon)=g(\epsilon)$, hence (10) is a symmetric Student’s-$t$ with $\nu$ 
degrees of freedom. The assumption that the error term follows a Student’s-$t$ is not a particularly strong 
hypothesis. It is a distribution that offers a good summary of the features of other (non-Gaussian) fat-tailed and 
symmetric distributions. Furthermore, our Monte Carlo experiments consider $N=1000$ replications, four 
different sample sizes, $T=(100, 200, 500, 1000)$, and the following combinations of causal and noncausal 
coefficients:
\begin{itemize}
\item MAR(1,1) :$\phi_{0}=0.8$, $\varphi_{0}=0.8$; time-reversible process;
\item MAR(1,1) :$\phi_{0}=0.8$, $\varphi_{0}=0.5$; time-irreversible process;
\item MAR(1,1) :$\phi_{0}=0.8$, $\varphi_{0}=0.1$; time-irreversible process;
\item MAR(1,0) :$\phi_{0}=0.8$; time-irreversible process.
\end{itemize}
In our Monte Carlo study, we also include results obtained by Ramsey and Rotham’s test, setting $k=2$.\\
\indent Table 1 shows the frequencies with which the two new strategies and the test proposed by Ramsey and 
Rotham detect the processes as time-irreversible when $\gamma=1$, $p$ is known, and $r$ and $s$ are 
unknown. In particular, columns Strategy 1 and Strategy 2 indicate the percentage of times the stochastic 
processes are identified as time-irreversible when the strategies from Sections 3.1 and 3.2 are 
implemented. The last column, $RR \ (1996)$, indicates how often we reject the null hypothesis of TR when 
the methodology proposed by \cite{ramsey1996time} is used. The Bayesian Information Criteria (BIC) is used
in Strategy 1. The results exhibit that Strategy 1 detects TR with greater precision, but is "undersized" for large 
$T$. This is because the penalty terms can differ from a tiny number in a large sample. On 
the other hand, Strategy 2 looks consistent and performs better when the processes under investigation are 
time-irreversible (frequencies are not size-adjusted, though, which makes the results of Strategies 1 and 2 not 
easy to compare). Finally, the test proposed by \cite{ramsey1996time} clearly shows size distortion problems. 
This is not an unexpected result since, as previously stated, the test can show a seriously distorted size if the 
distribution lacks a finite sixth moment. The Student’s-$t$ distribution has a finite sixth moment for
$\nu > 6$. As a consequence, the power of the test also performs poorly for RR (1996).\\
\indent Our simulation studies also consider cases where the error term is characterized by $\gamma \neq 1$. In 
these scenarios, we simulate a process with a skewed error term and proceed as if $\gamma=1$: Strategies 1 
and 2 are followed assuming a symmetric Student's-$t$ distributed error term. The results obtained under these 
new circumstances are similar to those in Table 1. This suggests that the test size and power are 
not sensitive to the eventual asymmetry of the error term. The outcomes are available upon request.\\
\indent Table 2 shows different results when $p$ is assumed unknown. In this case, before implementing our 
strategies, we estimate a pseudo-causal model in each replica of our simulation study to capture the dynamics 
$p$. Since there is more uncertainty under these new conditions, the results are less precise with small sample 
sizes ($T=(100, 200)$). However, the table displays that the outcomes align with Table 1 for large values of $T$. 
The percentages displayed in the column $RR(1996)$ of Table 2 are unchanged from those shown in Table 1 
since the same method is applied.
\begin{table}[H]
\centering
\scalebox{.55}{
\resizebox{\columnwidth}{!}{
\begin{tabular}{l c c c c c c c c c c c c c c c c c c}
\toprule 
\multicolumn{9}{c}{\ \ \ \ \ \ \ \ \ \ \ \ MAR(1,1); $\phi_{0}=0.8, \ \varphi_{0}=0.8, \ \nu_{0}=3$, $\gamma = 1$} \\
\toprule
&&& Strategy 1 &&& Strategy 2   &&&  RR (1996)\\
\toprule
T=100 	&&&7.1\%		&&&16.4\%	&&&8.1\%\\ 
T=200	&&&3.1\%	 	&&&7.5\%		&&&11.5\% \\
T=500	&&&1.4\%		&&&5.0\%		&&&11\% \\
T=1000 	&&&0.8\%		&&&4.5\%		&&&13.6 \%\\
\end{tabular}}}

\scalebox{.55}{
\resizebox{\columnwidth}{!}{
\begin{tabular}{l c c c c c c c c c c c c c c c c c c}
\toprule
\multicolumn{9}{c}{\ \ \ \ \ \ \ \ \ \ \ \ MAR(1,1); $\phi_{0}=0.8, \ \varphi_{0}=0.5, \ \nu_{0}=3$, $\gamma = 1$} \\
\toprule
&&& Strategy 1 &&& Strategy 2   &&&  RR (1996)\\
\toprule
T=100 	&&&51.4\%	&&&63.7\%	&&&20.7\%\\ 
T=200	&&&77.9\%	&&&84.8\%	&&&29.4\% \\
T=500	&&&99.0\%	&&&99.5\%	&&&40.7\% \\
T=1000 	&&&100\%	&&&100\%	&&&51.8 \%\\
\end{tabular}}}

\scalebox{.55}{
\resizebox{\columnwidth}{!}{
\begin{tabular}{l c c c c c c c c c c c c c c c c c c}
\toprule
\multicolumn{9}{c}{\ \ \ \ \ \ \ \ \ \ \ \ MAR(1,1); $\phi_{0}=0.8, \ \varphi_{0}=0.1, \ \nu_{0}=3$, $\gamma = 1$} \\
\toprule
&&& Strategy 1 &&& Strategy 2   &&&  RR (1996)\\
\toprule
T=100 	&&&87.4\%	&&&93.2\%		&&&33.2\%\\ 
T=200	&&&99.6\%	&&&99.9\%		&&&43.2\% \\
T=500	&&&100\%	&&&100\%		&&&57.5\% \\
T=1000 	&&&100\%	&&&100\%		&&&68.4\% \\
\end{tabular}}}

\scalebox{.55}{
\resizebox{\columnwidth}{!}{
\begin{tabular}{l c c c c c c c c c c c c c c c c c c}
\toprule
\multicolumn{9}{c}{\ \ \ \ \ \ \ \ \ \ \ \ MAR(1,0); $\phi_{0}=0.8, \ \nu_{0}=3$, $\gamma = 1$} \\
\toprule
&&& Strategy 1 &&& Strategy 2   &&&  RR (1996)\\
\toprule
T=100 	&&&91.5\%	&&&93.2\%		&&&34.2\%\\ 
T=200	&&&99.6\%	&&&99.9\%		&&&43.8\% \\
T=500	&&&100\%	&&&100\%		&&&58.9\% \\
T=1000 	&&&100\%	&&&100\%		&&&70.1\% \\
\bottomrule
\end{tabular}}}

\caption{\footnotesize{\textit{Frequencies with which time irreversibility is detected when the error term has
a symmetric Student's-$t$ distribution ($\gamma$=1) and $\nu_{0}=3$. Finally, $r$ and $s$ are assumed as unknown and $p$ as known.}}}
\end{table}

\begin{table}[H]
\centering
\scalebox{.55}{
\resizebox{\columnwidth}{!}{
\begin{tabular}{l c c c c c c c c c c c c c c c c c c}
\toprule 
\multicolumn{9}{c}{\ \ \ \ \ \ \ \ \ \ \ \ MAR(1,1); $\phi_{0}=0.8, \ \varphi_{0}=0.8, \ \nu_{0}=3$, $\gamma = 1$} \\
\toprule
&&& Strategy 1 &&& Strategy 2   &&&  RR (1996)\\
\toprule
T=100 	&&&20.9\%	&&&21.6\%	&&&8.1\%\\ 
T=200	&&&9.5\%		&&&12.6\%	&&&11.5\% \\
T=500	&&&3.5\%		&&&7.1\%		&&&11\% \\
T=1000 	&&&4.3\%		&&&7.8\%		&&&13.6 \%\\

\end{tabular}}}

\scalebox{.55}{
\resizebox{\columnwidth}{!}{
\begin{tabular}{l c c c c c c c c c c c c c c c c c c}
\toprule
\multicolumn{9}{c}{\ \ \ \ \ \ \ \ \ \ \ \ MAR(1,1); $\phi_{0}=0.8, \ \varphi_{0}=0.5, \ \nu_{0}=3$, $\gamma = 1$} \\
\toprule
&&& Strategy 1 &&& Strategy 2   &&&  RR (1996)\\
\toprule
T=100 	&&&61.6\%	&&&67.1\%	&&&20.7\%\\ 
T=200	&&&79.0\%	&&&85.2\%	&&&29.4\% \\
T=500	&&&99.0\%	&&&99.5\%	&&&40.7\% \\
T=1000 	&&&100\%	&&&100\%	&&&51.8 \%\\
\end{tabular}}}

\scalebox{.55}{
\resizebox{\columnwidth}{!}{
\begin{tabular}{l c c c c c c c c c c c c c c c c c c}
\toprule
\multicolumn{9}{c}{\ \ \ \ \ \ \ \ \ \ \ \ MAR(1,1); $\phi_{0}=0.8, \ \varphi_{0}=0.1, \ \nu_{0}=3$, $\gamma = 1$} \\
\toprule
&&& Strategy 1 &&& Strategy 2   &&&  RR (1996)\\
\toprule
T=100 	&&&92.2\%	&&&93.8\%	&&&33.2\%\\ 
T=200	&&&99.8\%	&&&99.9\%	&&&43.2\% \\
T=500	&&&100\%	&&&100\%	&&&57.5\% \\
T=1000 	&&&100\%	&&&100\%	&&&68.4\% \\
\end{tabular}}}

\scalebox{.55}{
\resizebox{\columnwidth}{!}{
\begin{tabular}{l c c c c c c c c c c c c c c c c c c}
\toprule
\multicolumn{9}{c}{\ \ \ \ \ \ \ \ \ \ \ \ MAR(1,1); $\phi_{0}=0.8, \ \nu_{0}=3$, $\gamma = 1$} \\
\toprule
&&& Strategy 1 &&& Strategy 2   &&&  RR (1996)\\
\toprule
T=100 	&&&95\%		&&&95.7\%	 &&&34.2\%\\ 
T=200	&&&99.9\%	&&&99.9\%	 &&&43.8\% \\
T=500	&&&100\%	&&&100\%	 &&&58.9\% \\
T=1000 	&&&100\%	&&&100\%	 &&&70.1\%\\
\bottomrule
\end{tabular}}}
\caption{\footnotesize{\textit{Frequencies with which time irreversibility is detected when the error term has a symmetric Student’s-t distribution ($\gamma$=1) with $\nu_{0} = 3$ and $p$, $r$, and $s$ are assumed as unknown.}}}
\end{table}
Finally, to analyze the result sensitivity to the persistence level, we implement new Monte Carlo experiments considering as $dgp$ a MAR(1,1) with the following combinations of causal and noncausal coefficients:
\begin{itemize}
\item MAR(1,1) :$\phi_{0}=0.95$, $\varphi_{0}=0.95$; time-reversible process;
\item MAR(1,1) :$\phi_{0}=0.95$, $\varphi_{0}=0.5$; time-irreversible process;
\item MAR(1,1) :$\phi_{0}=0.95$, $\varphi_{0}=0.1$; time-irreversible process;
\item MAR(1,0) :$\phi_{0}=0.95$; time-irreversible process.
\end{itemize}
Even in this case, the outcomes are similar to those displayed in Tables 1 and 2 and available upon request.

\section{Testing for time reversibility on non-stationary processes}
The goal of this section is to detect TR in non-stationary processes $\{ Y_{t} \}_{t=1}^{T} $ that can be 
expressed as:
\begin{equation}
	Y_{t}=f_{t}^{Y}+cc_{t}^{Y},
\end{equation}
where $f^{Y}$ is a generic trend function, and $cc^{Y}$ is a stationary process that captures the cyclical 
fluctuations of $Y$ around $f^{Y}$. We show that whenever the trend component is computed using the HP 
filter, then $f^{Y}$ can be expressed as a time-reversible process. As a consequence, the potential time 
irreversibility of process $Y$ would be captured by its cyclical component $cc^{Y}$. In other words, whenever 
$f^{Y}$ is estimated by using the HP filter, model (11) is time-irreversible (or reversible) if and only if its cyclical 
component is irreversible (or reversible).\\
\indent The HP filter estimates the trend component through the following minimization problem (see \cite{hecq21predicting}):
\begin{equation}
\begin{array}{rrclcl}
\displaystyle \min_{ \{ f_{t}^{Y} \}_{t=1}^{T} } & \Bigl( \sum_{t=1}^{T} y_{t}^{2}+\lambda \sum_{t=1}^{T} \bigl[ (f_{t}-f_{t-1}) - (f_{t-1}-f_{t-2}) \bigr]^{2}  \Bigr). \\
\end{array}
\end{equation}
\indent According to \cite{de2016econometrics}, the optimization problem (11) has the following closed-form 
solution:
\begin{equation}
	f_{t}^{Y} = \Bigl( \lambda L^{-2} - 4 \lambda L^{-1} + (1+6 \lambda) - 4\lambda L + \lambda^{2} \Bigr)^{-1} Y_{t},
\end{equation}
for $t=3, \dots, T-2$. The $\lambda$ parameter penalizes the filtered trend’s variability; therefore, the higher its 
value, the smoother the trend component:
\begin{equation*}
	\lambda=\Bigl( \frac{number \ of \ observations \ per \ year}{4} \Bigr)^{i} \times 1600,
\end{equation*}
with either $i = 2$ (see \cite{backus1992international}) or $i = 4$ (\cite{ravn2002adjusting}). It can be shown 
that (12) can be rewritten as:
\begin{equation}
	\begin{split}
		f_{t}^{Y} = \Biggl[ \Big( 1-\psi_{1}(\lambda) L - \psi_{2} (\lambda) L^{2} \Bigr) \Bigl( 1-\psi_{1}(\lambda) L^{-1} - \psi_{2} (\lambda) L^{-2} \Bigr) + \\
		- \Bigl( \psi_{1}^{2} (\lambda) + \psi_{2}^{2} (\lambda) + 6 \psi_{2} (\lambda) \Bigr) \Biggr]^{-1}Y_{t},
	\end{split}
\end{equation}
where $\psi_{1}(\lambda)=\frac{4 \lambda}{\lambda+1}$, and $\psi_{2} (\lambda ) = - \lambda$. For instance, 
for annual data, we can adopt $\lambda=6.25$, implying
\begin{equation*}
	f_{t}^{Y} = \Biggl[ \Big( 1-\frac{100}{29} L + 6.25 L^{2} \Bigr) \Bigl( 1- \frac{100}{29} L^{-1} + 6.25 L^{-2} \Bigr) - 13.456 \Biggr]^{-1}Y_{t}.
\end{equation*}
The results underline that the filter of the trend component is given by a time-reversible MAR(2,2) polynomial 
minus a constant value. Since the latter does not affect the symmetry over time of our process, and our goal is 
to investigate the time reversibility of $f^{Y}$, we do not consider the constant term in our investigation. As a 
consequence, we can approximate $f^{Y}$ as follows:
\begin{equation}
	f_{t}^{Y} \approx \Biggl[ \Big( 1-\frac{100}{29} L + 6.25 L^{2} \Bigr) \Bigl( 1- \frac{100}{29} L^{-1} + 6.25 L^{-2} \Bigr) \Biggr]^{-1}Y_{t}.
\end{equation}
Using the Laurent expansion as in (4), we have:
\begin{equation} 
	f_{t}^{Y} \approx \sum_{j= - \infty}^{+ \infty} \delta _{j} Y_{t-j},
\end{equation}
where because of the identity of the lead and lag polynomials, $\delta$ is symmetric over time. Hence, even if 
$f^{Y}$ is a non-stationary process, we can apply a weaker definition of TR and define it as time-reversible. 
This result implies that the potential time irreversibility (or reversibility) lies with the cyclical component of $Y$.\\
\indent To illustrate how our new strategies perform under the new conditions, we implement new Monte Carlo experiments where $f^{Y}$ is a random walk with drift:
\begin{equation}
	X_{t}=X_{t-1}+ \delta + \eta_{t},
\end{equation}
with $\eta \sim N(0,1)$, and $cc^{Y}$ as a MAR(1,1):
\begin{equation}
	 (1+\phi L) (1+\varphi L^{-1}) \tilde{Y}_{t} =\varepsilon_{t}.
\end{equation}
Finally, the process $\{ Y \}_{t=1}^{T}$ is obtained by the sum of the two processes, that is:
\begin{equation}
	Y_{t}= X_{t}+ \tilde{Y}_{t}.
\end{equation}
Alternatively, we could have considered the following process as $dgp$:
\begin{equation}
	Y_{t}=Y_{t-1} + \delta + \tilde{Y}_{t} \Rightarrow \Delta Y_{t} = \delta + \tilde{Y}_{t}.
\end{equation}
However, the reason not to consider such a process is that, as expressed in (20), the resulting $dgp$ 
implies that the first difference process ($\Delta Y_{t}$) is a MAR(1,1). This is not a realistic assumption 
because MARs are typically used to capture explosive bubbles, and the first difference operation eliminates 
most locally explosive behaviors (see \cite{hecq21predicting}).\\
\indent In each replica of our Monte Carlo experiment, we simulate the non-stationary process $\{ Y_{t} \}_{t=1}^{T} $, remove the trend component using the HP filter, and then apply our strategies on $cc^{Y}$. 
The coefficients used for the cyclical component $cc^{Y}$ are the same as in the previous section. 
Table 3 displays the results.
\begin{table}[H]
	\centering
	\scalebox{.6}{
		\resizebox{\columnwidth}{!}{
			\begin{tabular}{l c c c c c c c c c c c c c c c c c c}
			\toprule 
			\multicolumn{9}{c}{\ \ \ \ \ \ \ \ $cc^{Y}: \ \ MAR(1,1); \ \phi_{0}=0.8, \ \varphi_{0}=0.8; \ \nu_{0}=3$, \ $\gamma=1$} \\
			\toprule
			&&& Strategy 1 &&& Strategy 2  \\
			\toprule
			T=100 	&&&15.1\%	&&&33.3\%\\ 
			T=200	&&&6.7\%	 	&&&8.1\%		\\
			T=500	&&&1.4\%		&&&5.0\%		\\
			T=1000 	&&&0.8\%		&&&4.5\%		\\
		\end{tabular}
	}
}

\scalebox{.6}{
\resizebox{\columnwidth}{!}{
\begin{tabular}{l c c c c c c c c c c c c c c c c c c}
\toprule
\multicolumn{9}{c}{\ \ \ \ \ \ \ \ $cc^{Y}: \ \ MAR(1,1); \ \phi_{0}=0.8, \ \varphi_{0}=0.5; \ \nu_{0}=3$, \ $\gamma=1$} \\
\toprule
&&& Strategy 1 &&& Strategy 2   \\
\toprule
T=100 	&&&40.6\%	&&&59.5\%	\\ 
T=200	&&&58.0\% 	&&&69.7\%	 \\
T=500	&&&86.7\%	&&&94.8\%	\\
T=1000 	&&&99.2\%	&&&100\%	\\
\end{tabular}}}

\scalebox{.6}{
\resizebox{\columnwidth}{!}{
\begin{tabular}{l c c c c c c c c c c c c c c c c c c}
\toprule
\multicolumn{9}{c}{\ \ \ \ \ \ \ \ $cc^{Y}: \ \ MAR(1,1); \ \phi_{0}=0.8, \ \varphi_{0}=0.1; \ \nu_{0}=3$, \ $\gamma=1$} \\
\toprule
&&& Strategy 1 &&& Strategy 2   \\
\toprule
T=100 	&&&71.9\%	&&&89.2\%\\ 
T=200	&&&92.0\%	&&&97.3\% \\
T=500	&&&99.6\%	&&&100\% \\
T=1000 	&&&100\%	&&&100\% \\
\end{tabular}}}

\scalebox{.6}{
\resizebox{\columnwidth}{!}{
\begin{tabular}{l c c c c c c c c c c c c c c c c c c}
\toprule
\multicolumn{9}{c}{\ \ \ \ \ \ \ \ \ \ \ $cc^{Y}: \ \ \ \ \ MAR(1,0); \ \ \phi_{0}=0.8; \ \  \nu_{0}=3$, \ \ $\gamma=1$} \ \ \ \ \ \ \ \\
\toprule
&&& Strategy 1 &&& Strategy 2   \\
\toprule
T=100 	&&&75.1\%	&&&91.1\%\\ 
T=200	&&&93.4\%	&&&98.8\% \\
T=500	&&&100\%	&&&100\% \\
T=1000 	&&&100\%	&&&100\% \\
\bottomrule
\end{tabular}}}
\caption{\footnotesize{\textit{Frequencies with which time irreversibility is detected on non-stationary time series; 
$r$ and $s$ are assumed as unknown and $p$ as known. Data are considered to have quarterly frequency 
($\lambda=1600$).}}}
\end{table}
The results are similar to those displayed in Tables 1 and 2, with the difference that the power of the strategies is less accurate under these new conditions, especially when the sample size considered is small ($T=(100,200)$).

\section{Is climate change time-reversible?}
In our empirical investigation, we analyze annual data for the global land and ocean temperature anomaly 
($GLO$), the global land temperature anomaly ($GL$), the global ocean temperature anomaly ($GO$), solar 
activity ($SA$), emissions of greenhouses gas ($GHG$), emissions of nitrous oxide ($N2O$). When available, 
we also use monthly data to control for potential small sample distortions in our statistics, as revealed by the 
simulation results. In particular, we consider the following monthly series: the Southern Oscillation Index 
($SOI$), the North Atlantic Oscillation Index ($NAO$), the Pacific Decadal Oscillation Index ($PDO$), the 
global mean sea level ($GMSL$), the Northern Hemisphere sea ice area ($NH$), the Southern Hemisphere 
sea ice area ($SH$), the global component of climate at a glance ($GCAG$), and, finally, the global surface 
temperature change ($GISTEMP$).\footnote{\tiny{$GLO$, $GL$, and $GO$ are obtained from https://www.ncdc.noaa.gov/cag/global/time-series. $SOI$ and $NAO$ are obtained from https://www.cpc.ncep.noaa.gov/data/indices/soi and \\ https://www.cpc.ncep.noaa.gov/products/precip/CWlink/pna/norm.nao.monthly.b5001.current.ascii.table, respectively. $PDO$ is obtained from https://www.ncdc.noaa.gov/teleconnections/pdo/. For $GHG$ and $SA$, the source is \cite{hansen2017young}. For $N2O$, we use the historical reconstruction computed in \cite{meinshausen2017historical} (data available at https://www.climatecollege.unimelb.edu.au/cmip). $NH$ and $SH$ are obtained from https://psl.noaa.gov/data/timeseries/monthly, $GCAG$ and $GISTEMP$ from https://datahub.io/core/global-temp and, finally, $GMSL$
from https://datahub.io/core/sea-level-rise.}} \\
\indent $GLO$, $GL$, $GO$, $GCAG$, and $GISTEMP$ measure global warming. They provide
the difference between the current temperature from a standard benchmark value. Positive anomalies 
show that the observed temperature is warmer than the benchmark value, and negative temperatures show 
that the observed temperature is colder than the benchmark value. In particular, $GCAG$ provides global-scale 
temperature information using data from NOAA’s Merged Land Ocean Global Surface Temperature 
Analysis (NOAAGlobalTemp), which uses comprehensive data collections of increased global coverage over 
land (Global Historical Climatology Network-Monthly) and ocean (Extended Reconstructed Sea Surface 
Temperature) surfaces.\\
\indent $SOI$ is one of the most important atmospheric indices for determining the strength of \textit{El Ni{\~n}o} 
and \textit{La Ni{\~n}a} events and their possible effects on weather conditions in the tropics and various other 
geographical areas. \textit{El Ni{\~n}o events} are characterized by sustained warming of the central and 
eastern tropical Pacific, whereas \textit{La Ni{\~n}a} events show sustained cooling of the same areas. 
These changes in the Pacific Ocean and its overlying atmosphere occur in a cycle known as the 
\textit{El Ni{\~n} o–Southern Oscillation} ($ENSO$). High values of $SOI$ indicate \textit{La Ni{\~n}a} events, 
whereas negative values indicate \textit{El Ni{\~n}o} events. The $NAO$ determines the westerly winds' speed 
and direction across the North Atlantic and the winter sea surface temperature. When the $NAO$ index is far 
above average, there is a greater likelihood that seasonal temperatures in northern Europe, northern Asia, 
and South-East North America will be warmer than usual. In contrast, seasonal temperatures in North Africa, 
North-East Canada, and southern Greenland will be cooler than usual. The opposite is true when $NAO$ is 
far below average. $PDO$ is a climatic cycle that describes anomalies in sea surface temperature in the 
Northeast Pacific Ocean. The $PDO$ has the power to influence weather patterns all over North America.  
Finally, $GMSL$, $NH$, and $SH$ are climate indicators providing information on how much of the ice land is
melting, and their connection with global warming is straightforward.\\
\indent Figure 1 presents the data. $GLO$, $GL$, $GO$, $SA$, $GHG$, and $N2O$ range from 1881 to 2014, $SOI$ and $NAO$ from January 1951 to December 2021, $PDO$ from January 1854 to December 2021, $GCAG$ and $GISTEMP$ from January 1880 to December 2016, $GMSL$ from January 1880 to December 2015, and, finally, $NH$ and $SH$ from January 1979 to 2021.
\begin{figure}[H]
	\begin{center}
 		 \begin{subfigure}[t]{.22\textwidth}
   		 	\centering
  		 	\includegraphics[width=\linewidth]{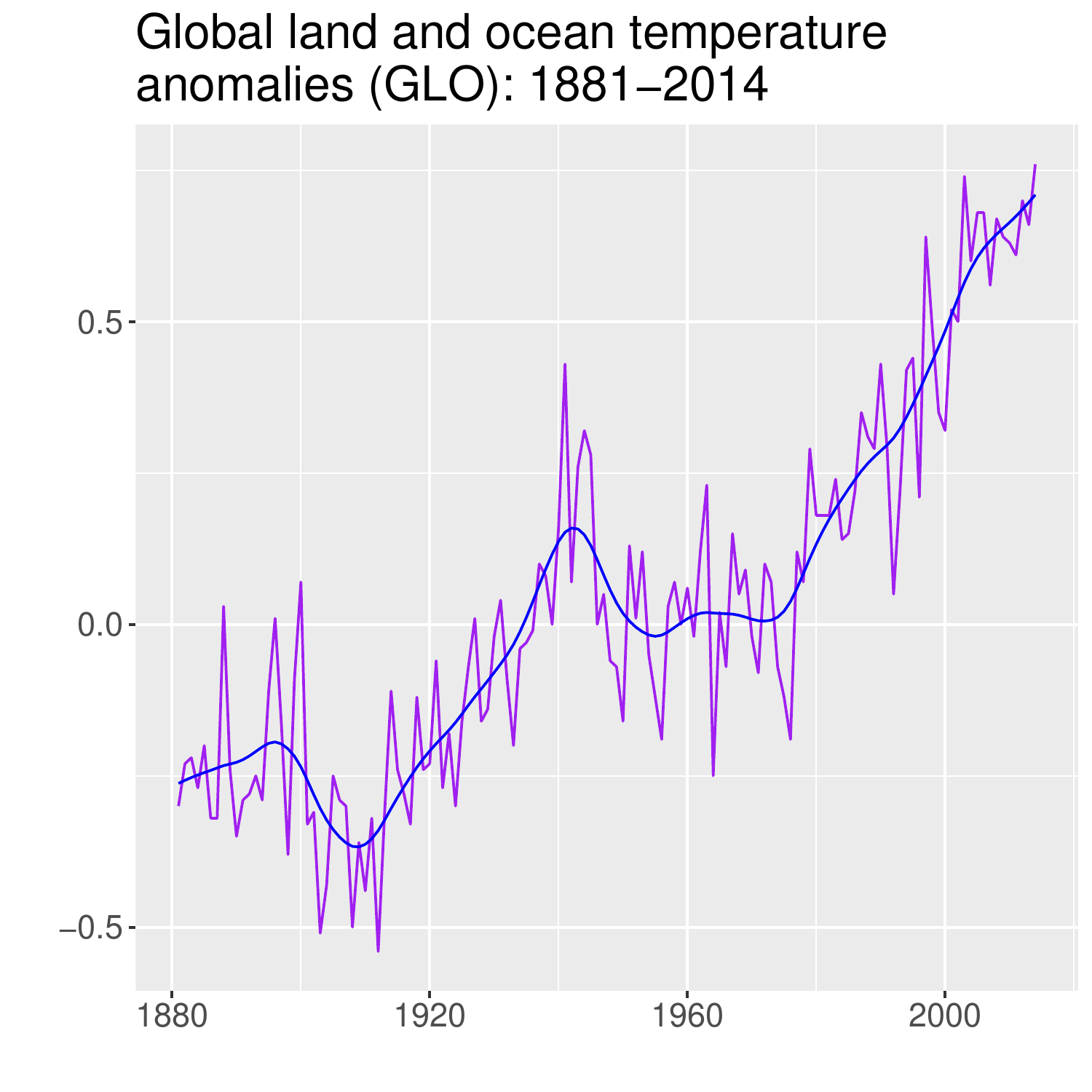}
 		  	\caption{\textit{Annual data for \\ global land and \\ ocean temperature \\ anomalies.}}
 		 \end{subfigure}
		\begin{subfigure}[t]{.22\textwidth}
   		 	\centering
   			 \includegraphics[width=\linewidth]{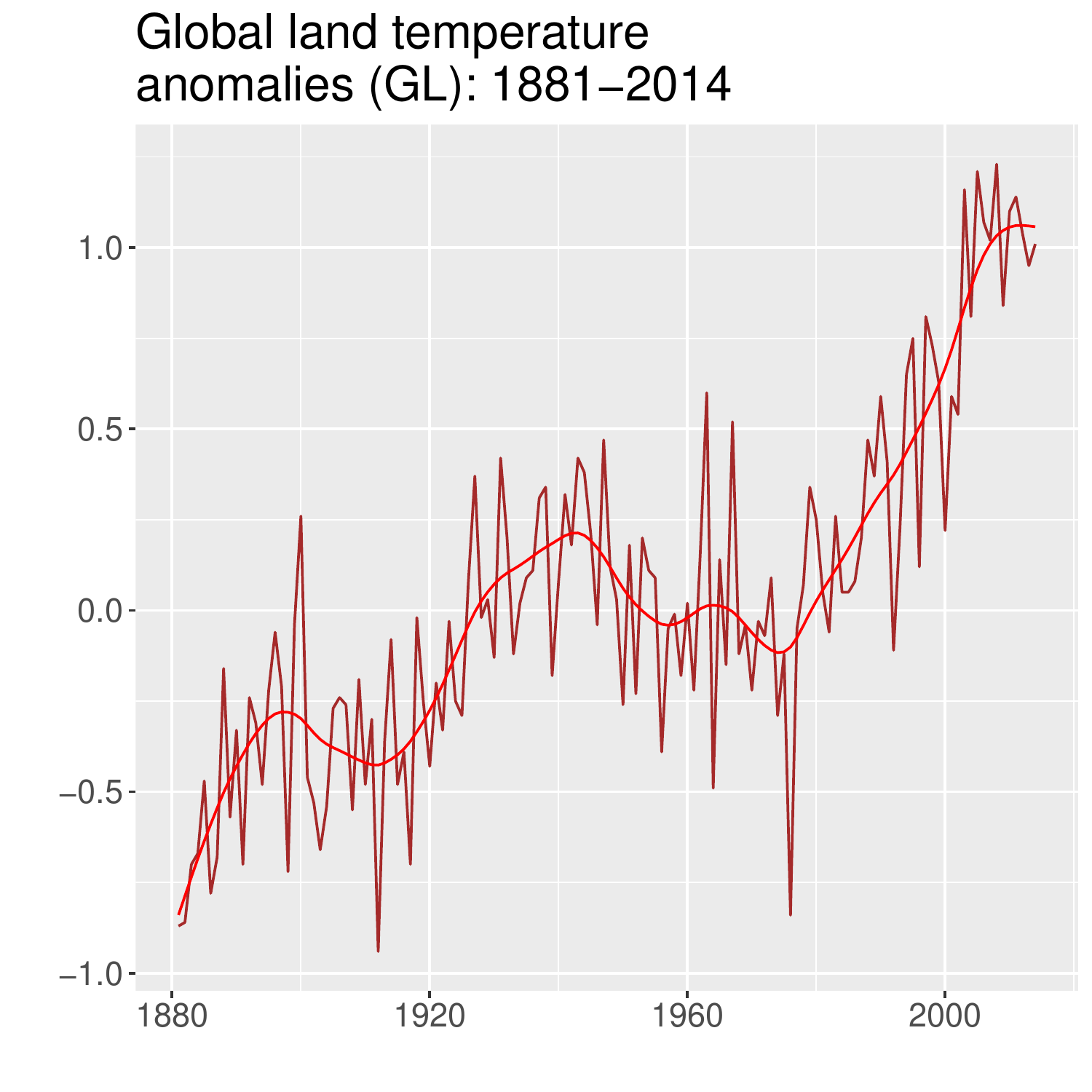}
  		 	\caption{\textit{Annual data for \\ global land \\ temperature \\ anomalies.}}
 		\end{subfigure}
		\begin{subfigure}[t]{.22\textwidth}
   		 	\centering
   			 \includegraphics[width=\linewidth]{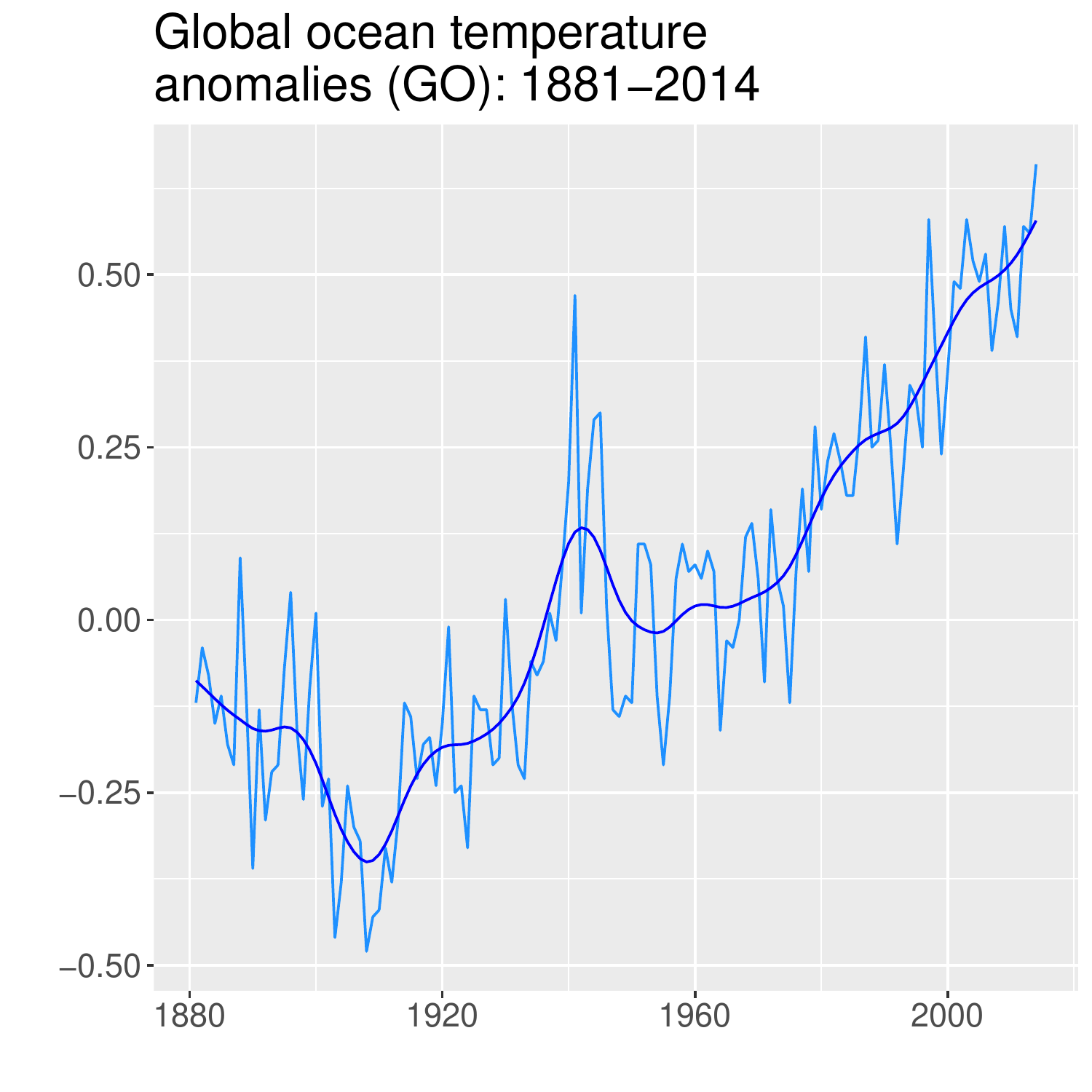}
  		 	\caption{\textit{Annual data for \\ global ocean \\ temperature \\ anomalies.}}
 		\end{subfigure}
		\begin{subfigure}[t]{.22\textwidth}
   		 	\centering
   			 \includegraphics[width=\linewidth]{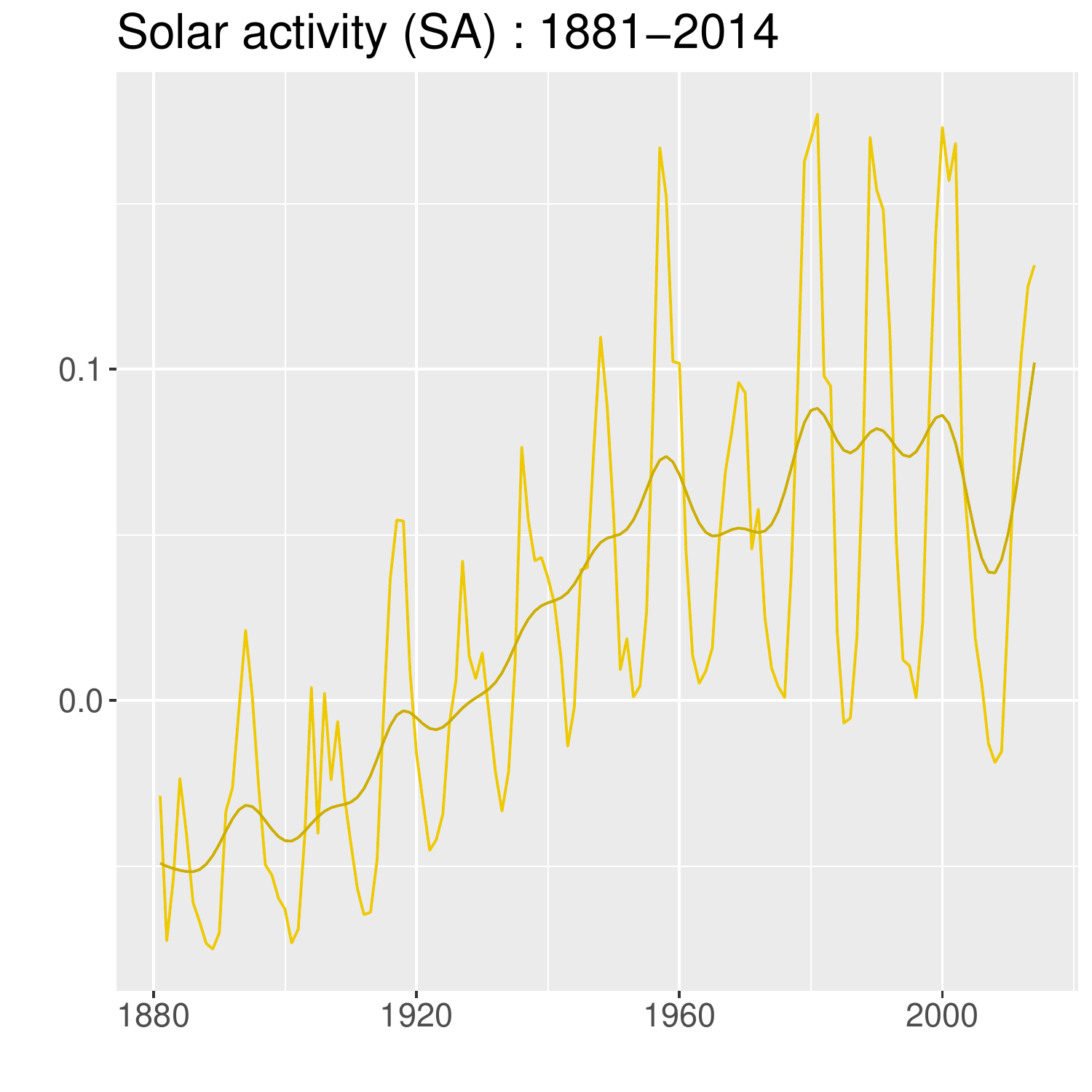}
  		 	\caption{\textit{Annual data for \\ solar activity.}}
 		\end{subfigure}
		
		\bigskip
	
		\begin{subfigure}[t]{.22\textwidth}
   		 	\centering
   			 \includegraphics[width=\linewidth]{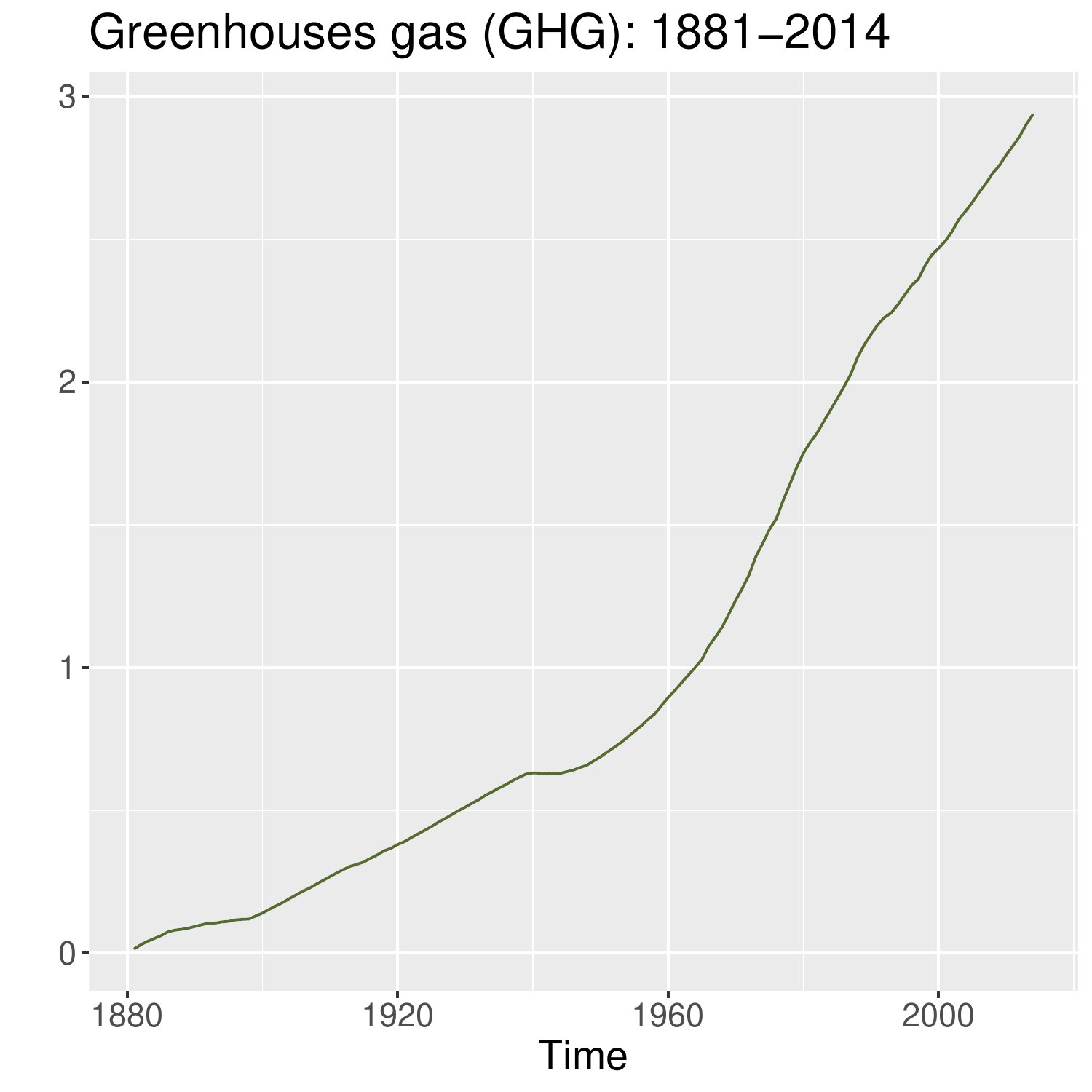}
  		 	\caption{\textit{Annual data for \\ greenhouses gas.}}
 		\end{subfigure}
		\begin{subfigure}[t]{.22\textwidth}
   		 	\centering
   			 \includegraphics[width=\linewidth]{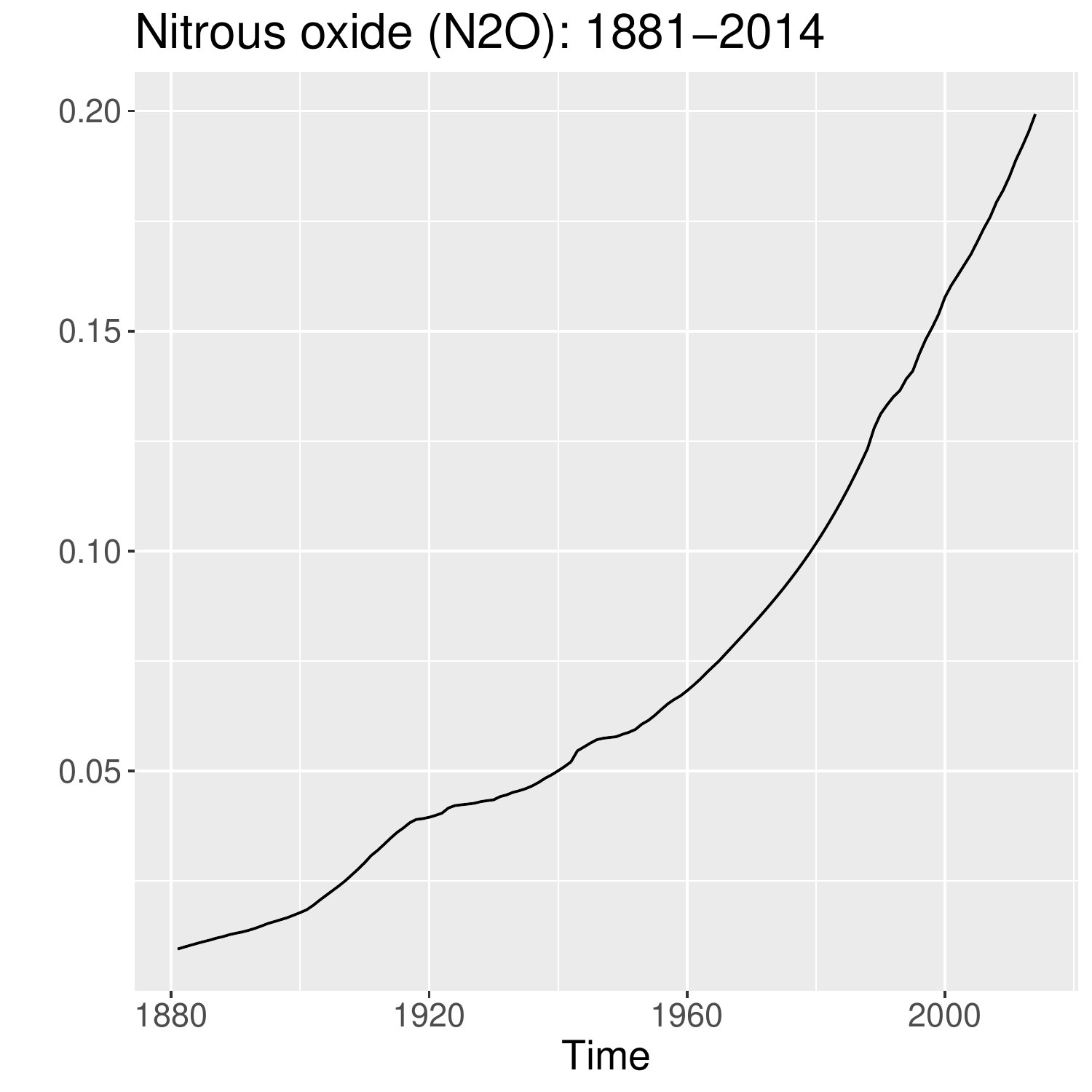}
  		 	\caption{\textit{Annual data for \\ nitrous oxide.}}
 		\end{subfigure}
		\begin{subfigure}[t]{.22\textwidth}
   		 	\centering
   			 \includegraphics[width=\linewidth]{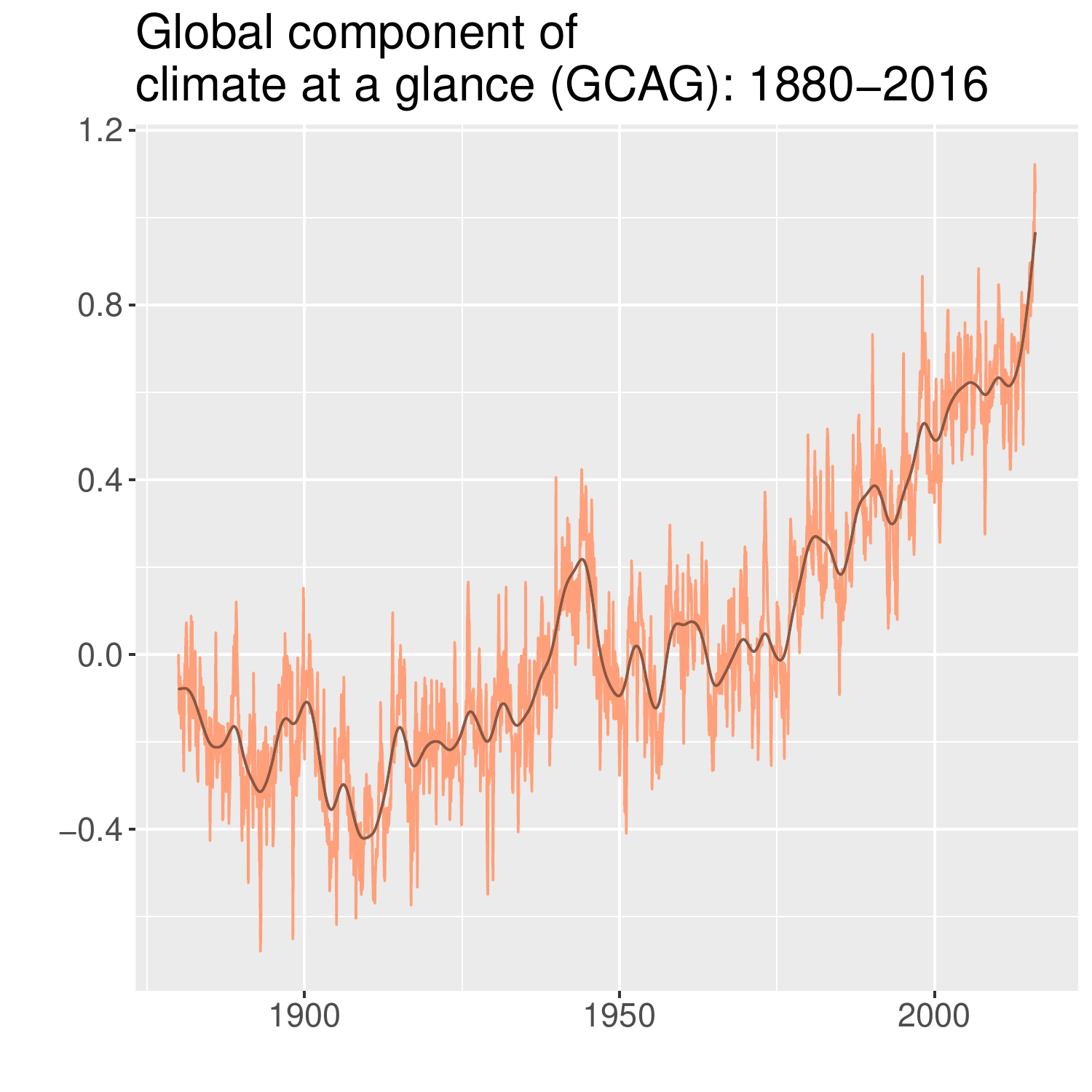}
  		 	\caption{\textit{Monthly data for \\ global component \\ of climate at \\ a glance.}}
 		\end{subfigure}
		\begin{subfigure}[t]{.22\textwidth}
   		 	\centering
   			 \includegraphics[width=\linewidth]{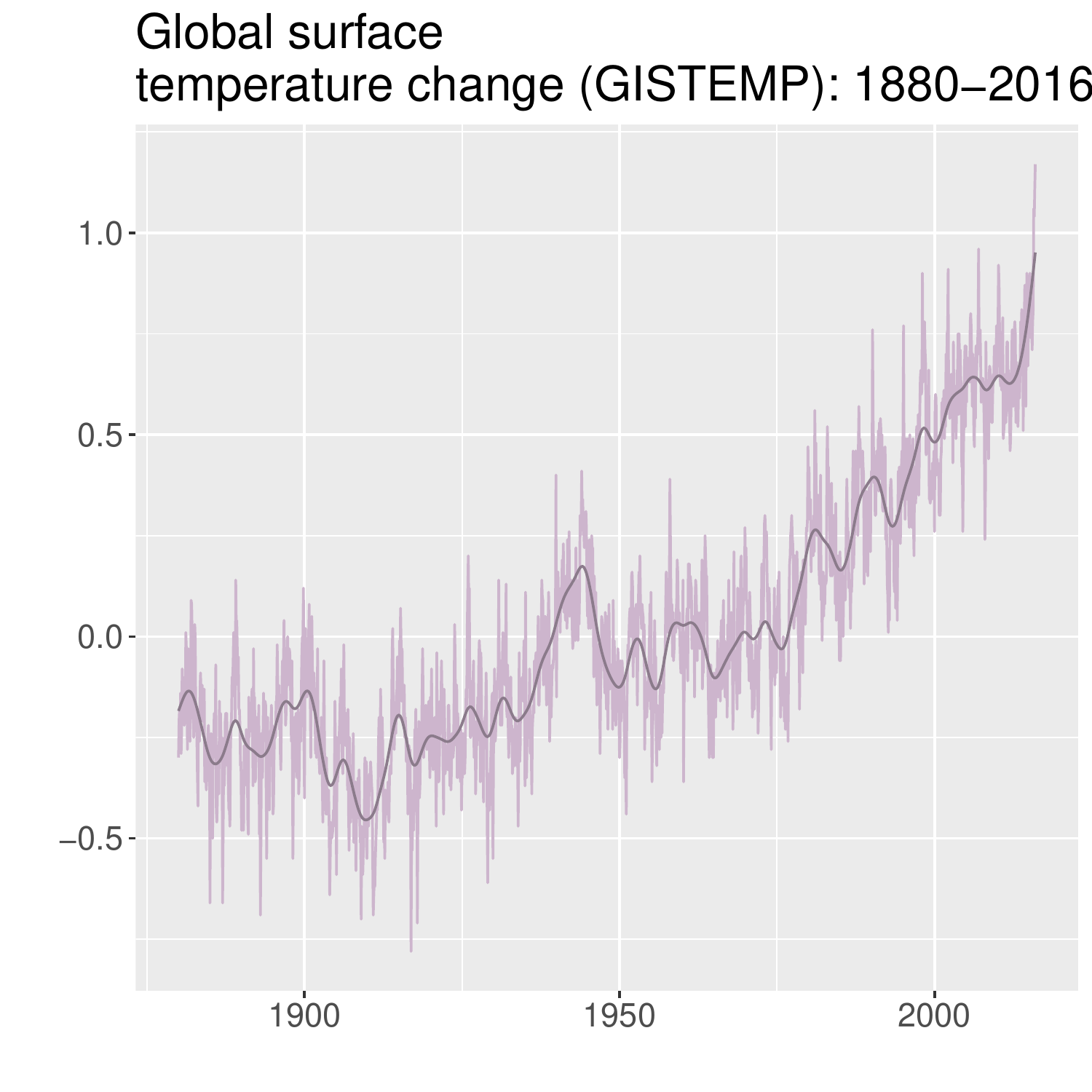}
  		 	\caption{\textit{Monthly data \\ for global surface \\ temperature change.}}
 		\end{subfigure}

		\bigskip
		
		\begin{subfigure}[t]{.22\textwidth}
   		 	\centering
   			 \includegraphics[width=\linewidth]{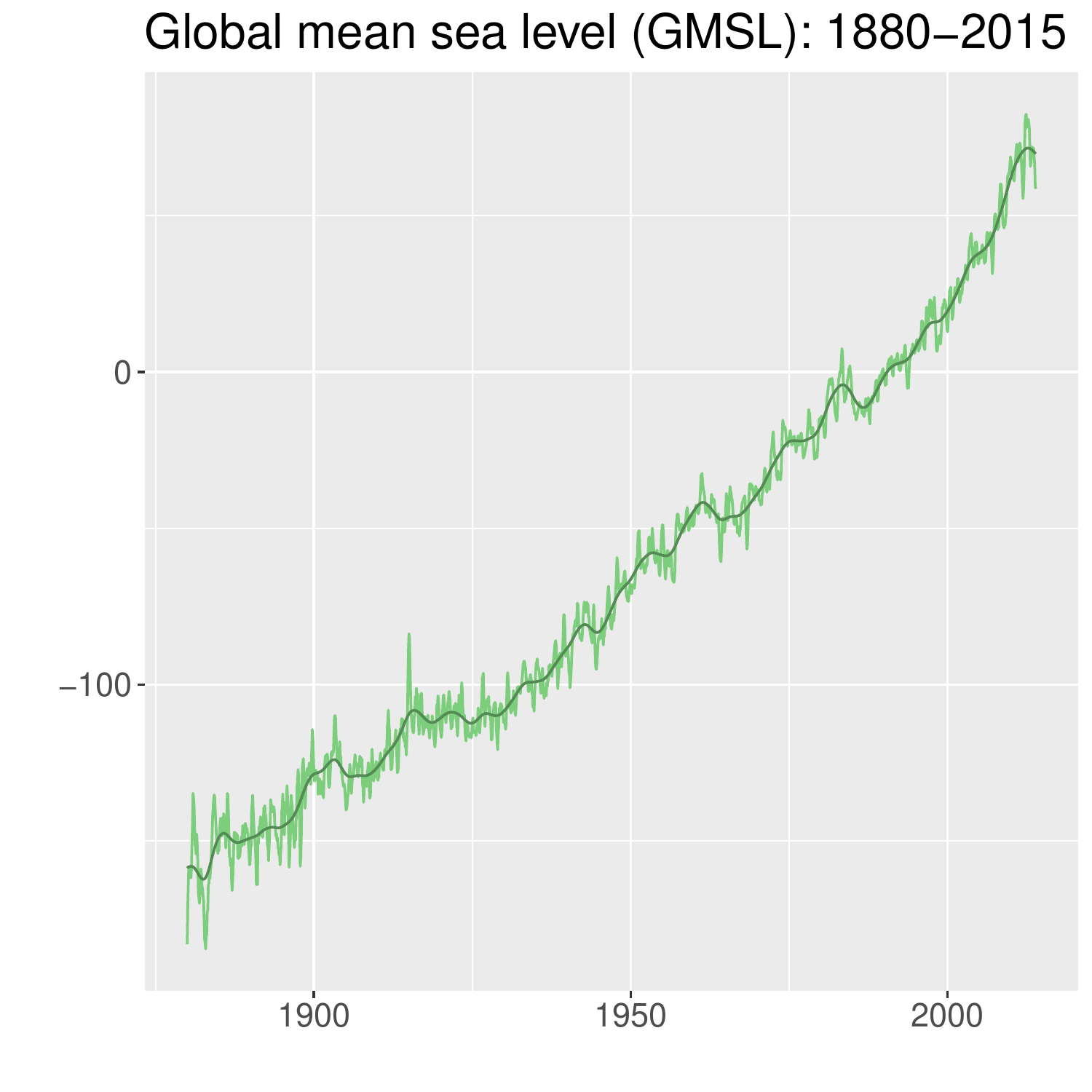}
  		 	\caption{\textit{Monthly data \\ for global mean \\ sea level.}}
 		\end{subfigure}		
		\begin{subfigure}[t]{.22\textwidth}
   		 	\centering
   			 \includegraphics[width=\linewidth]{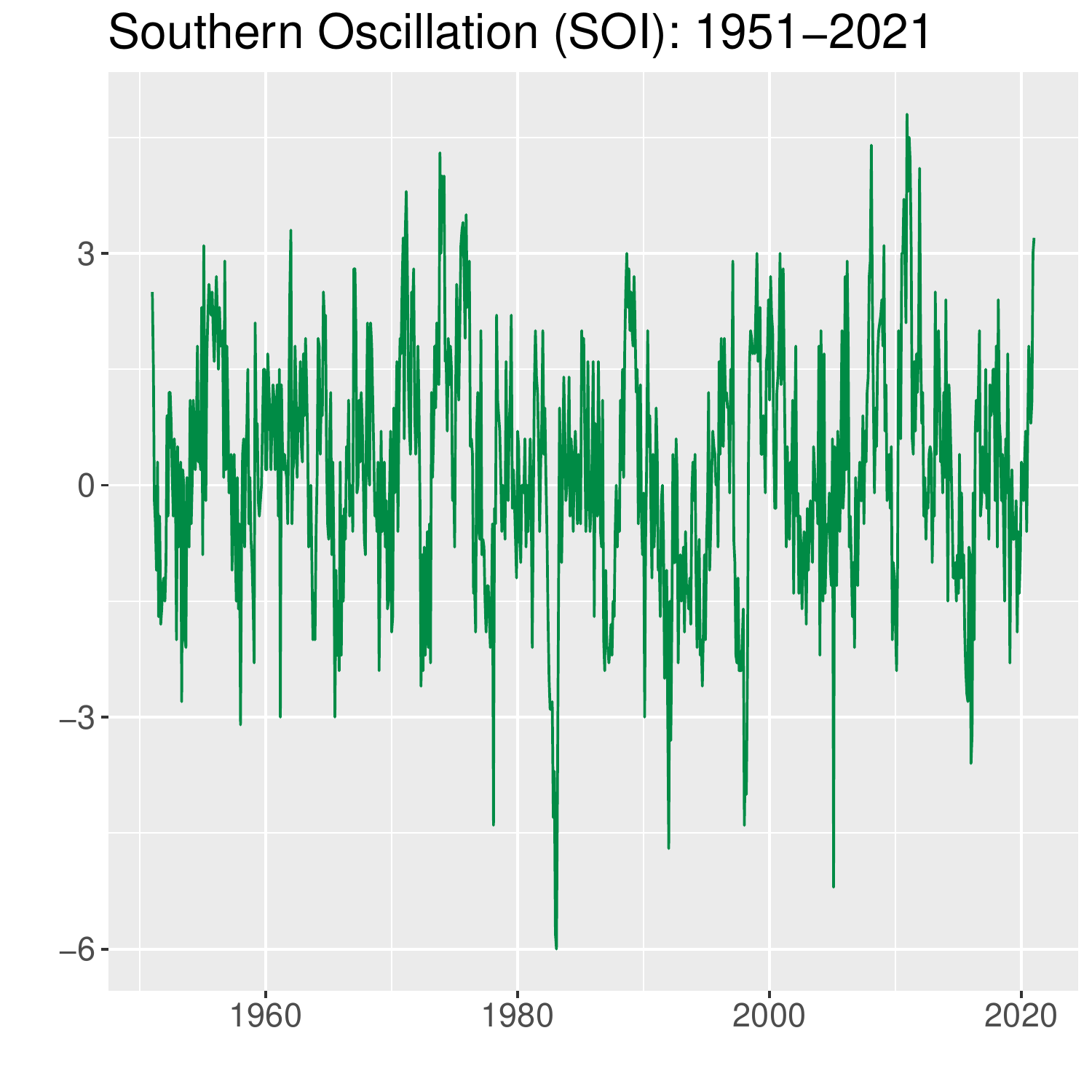}
  		 	\caption{\textit{Monthly data \\ for Southern \\ Oscillation index.}}
 		\end{subfigure}
		\begin{subfigure}[t]{.22\textwidth}
   		 	\centering
   			 \includegraphics[width=\linewidth]{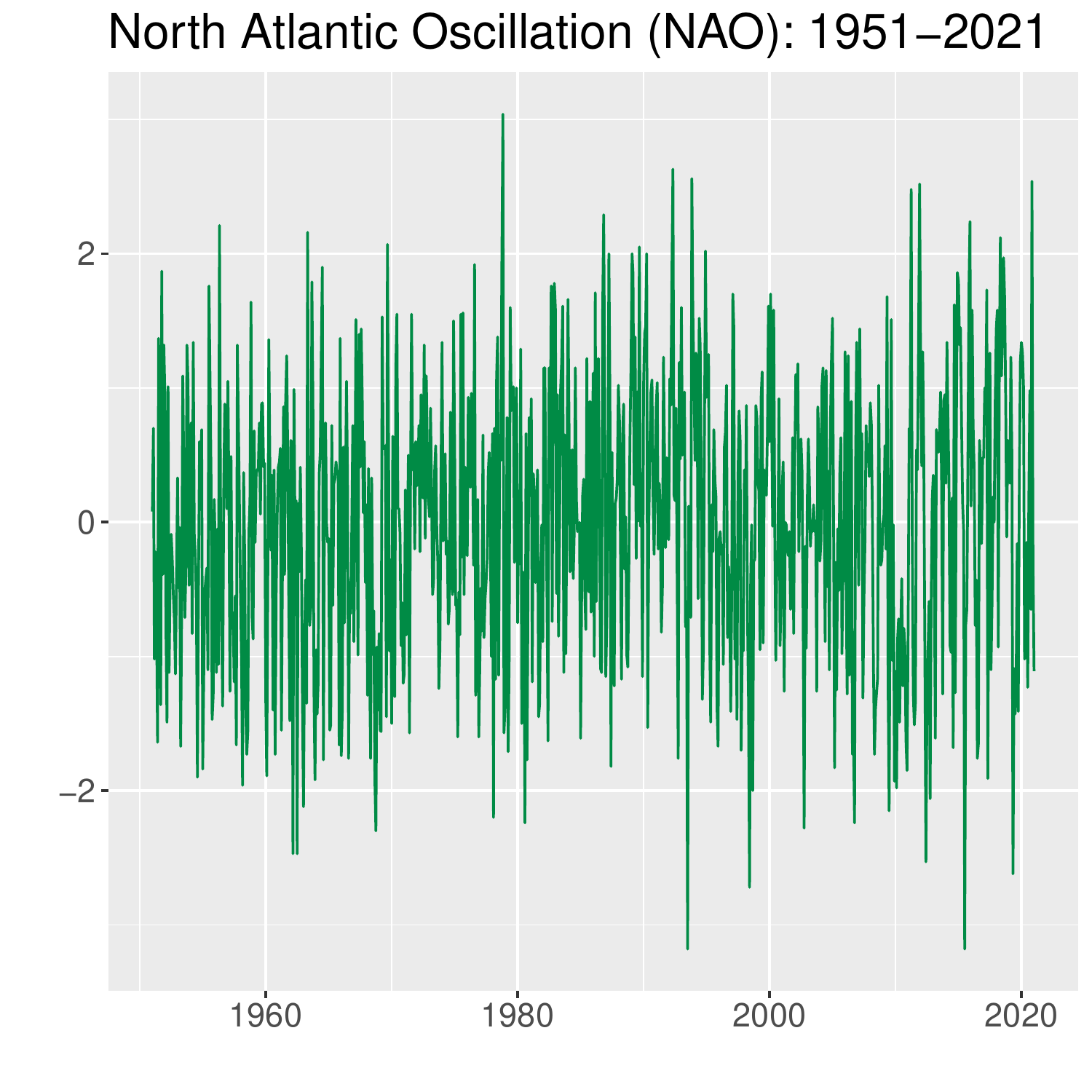}
  		 	\caption{\textit{Monthly data \\ for North Atlantic \\ Oscillation index.}}
 		\end{subfigure}
		\begin{subfigure}[t]{.22\textwidth}
   		 	\centering
   			 \includegraphics[width=\linewidth]{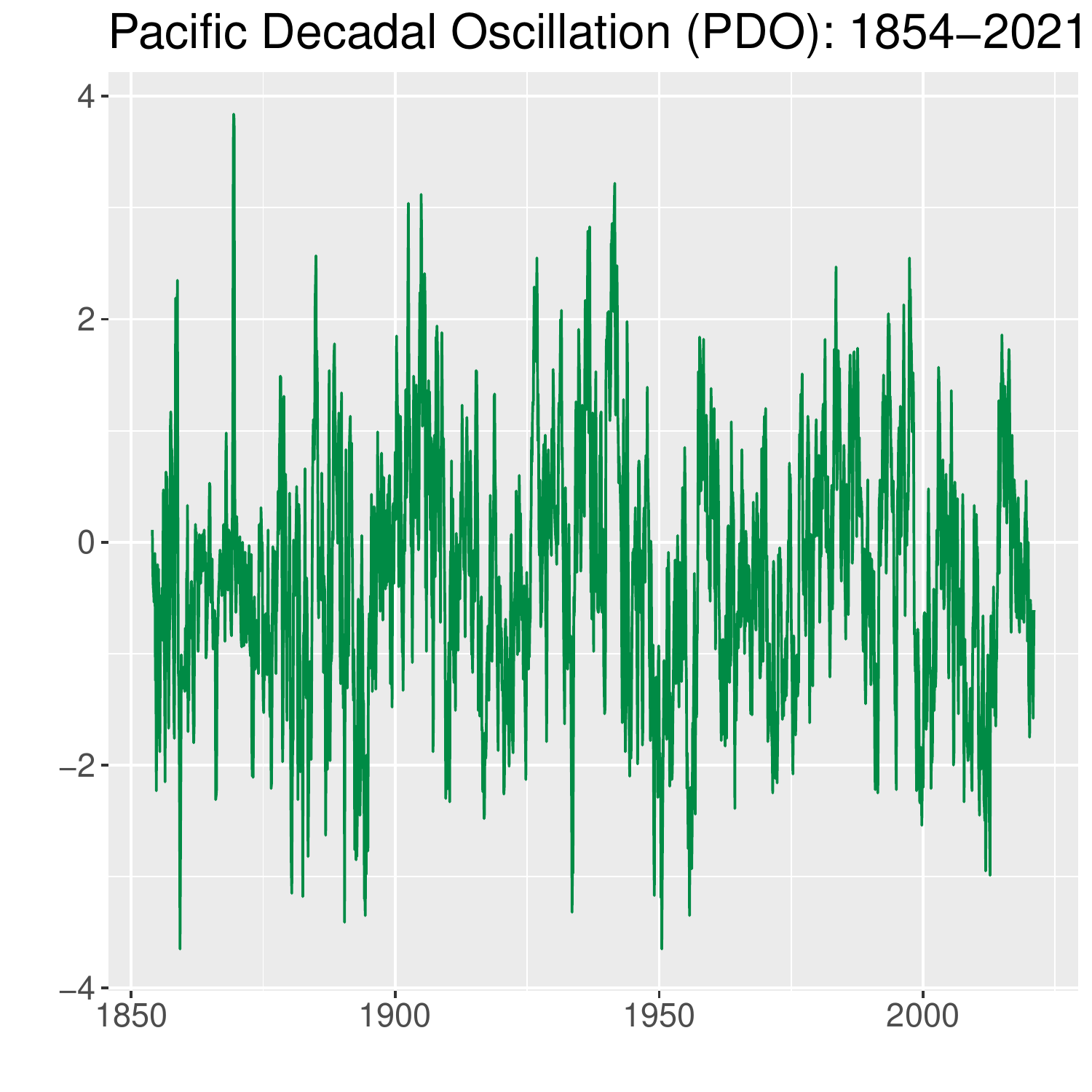}
  		 	\caption{\textit{Monthly data \\ for Pacific Decadal \\ Oscillation index.}}
 		\end{subfigure}
		
		\bigskip
		
		\begin{subfigure}[t]{.22\textwidth}
   		 	\centering
   			 \includegraphics[width=\linewidth]{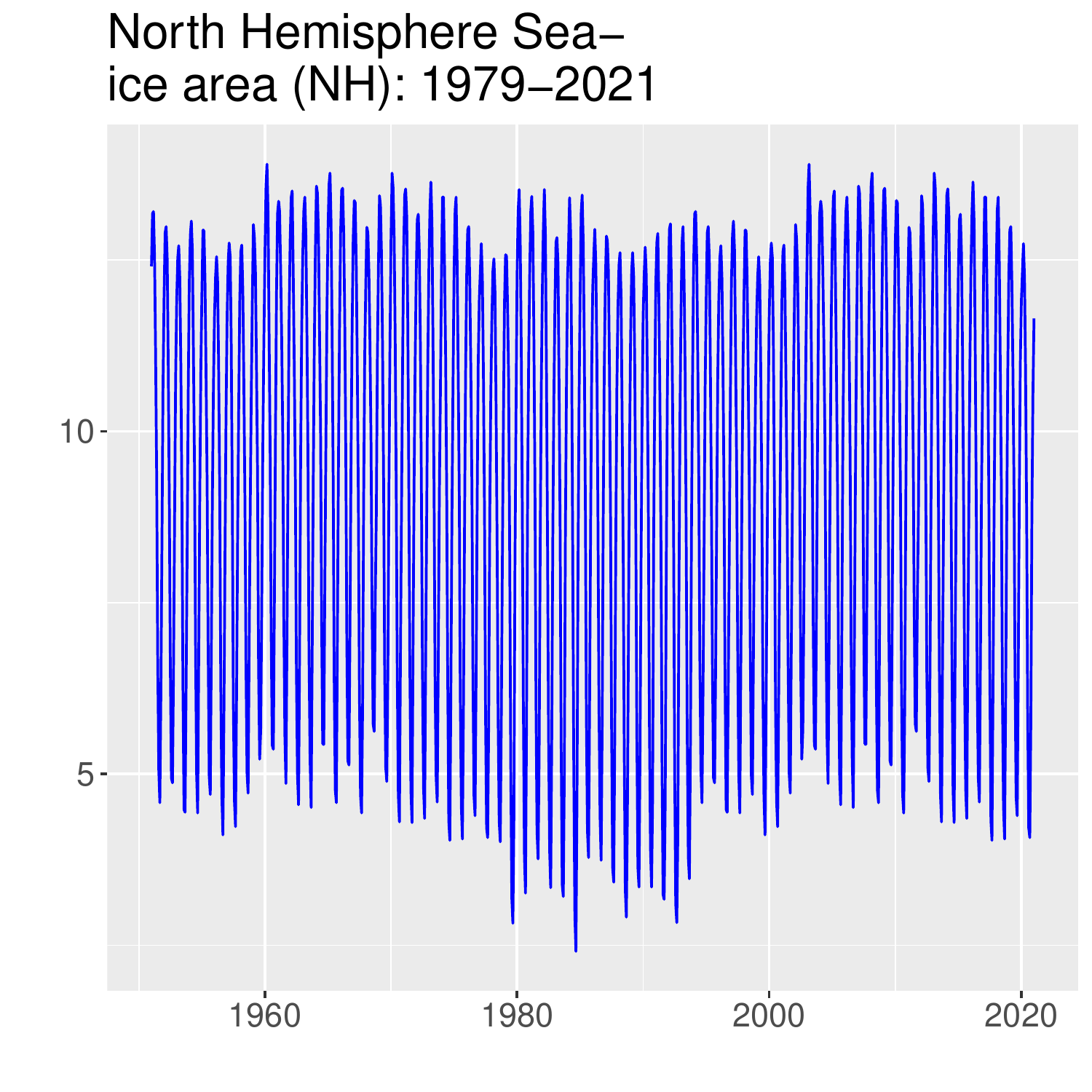}
  		 	\caption{\textit{Monthly data \\ for Northern \\ Hemisphere \\ sea ice area.}}
 		\end{subfigure}
		\begin{subfigure}[t]{.22\textwidth}
   		 	\centering
   			 \includegraphics[width=\linewidth]{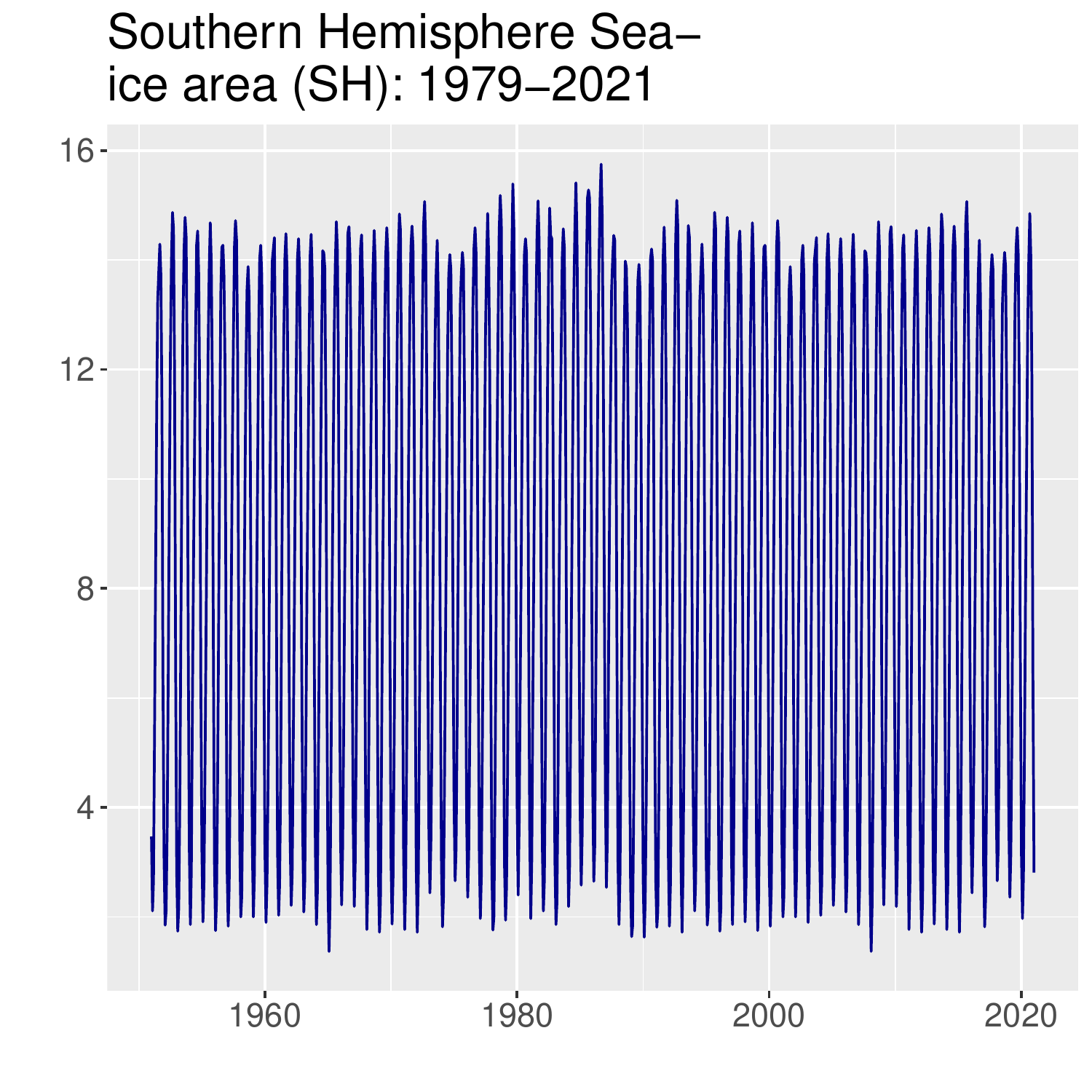}
  		 	\caption{\textit{Monthly data \\ for Southern \\ Hemisphere \\ sea ice area.}}
 		\end{subfigure}
	\end{center}
 	\caption{Climate time series.}
 \end{figure}
 
 \

\indent As shown in Figure 1, time series (a)-(i) are characterized by a positive trend. Hence, according to the strategy introduced in Section 4, their potential time-reversibility (or irreversibility) lies with their cyclical component. For this reason, we can remove their trend and extract their cyclical fluctuations using the HP filter. Figure 2 displays the detrended time series.
 \begin{figure}[H]
	\begin{center}
 		 \begin{subfigure}[t]{.24\textwidth}
   		 	\centering
  		 	\includegraphics[width=\linewidth]{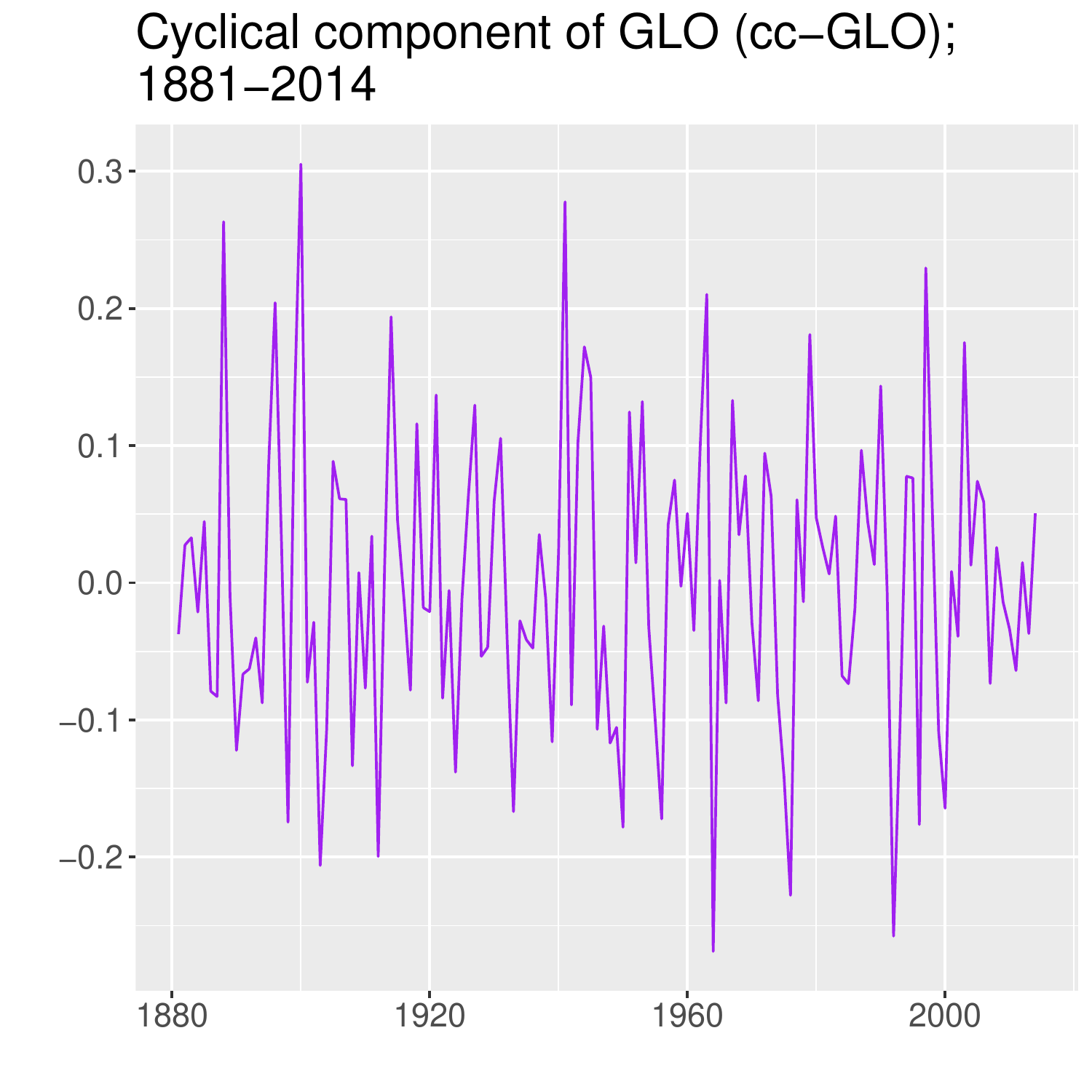}
 		  	\caption{\textit{Annual data for \\ cyclical component \\ of global land and \\ ocean temperature \\ anomalies.}}
 		 \end{subfigure}
		\begin{subfigure}[t]{.24\textwidth}
   		 	\centering
   			 \includegraphics[width=\linewidth]{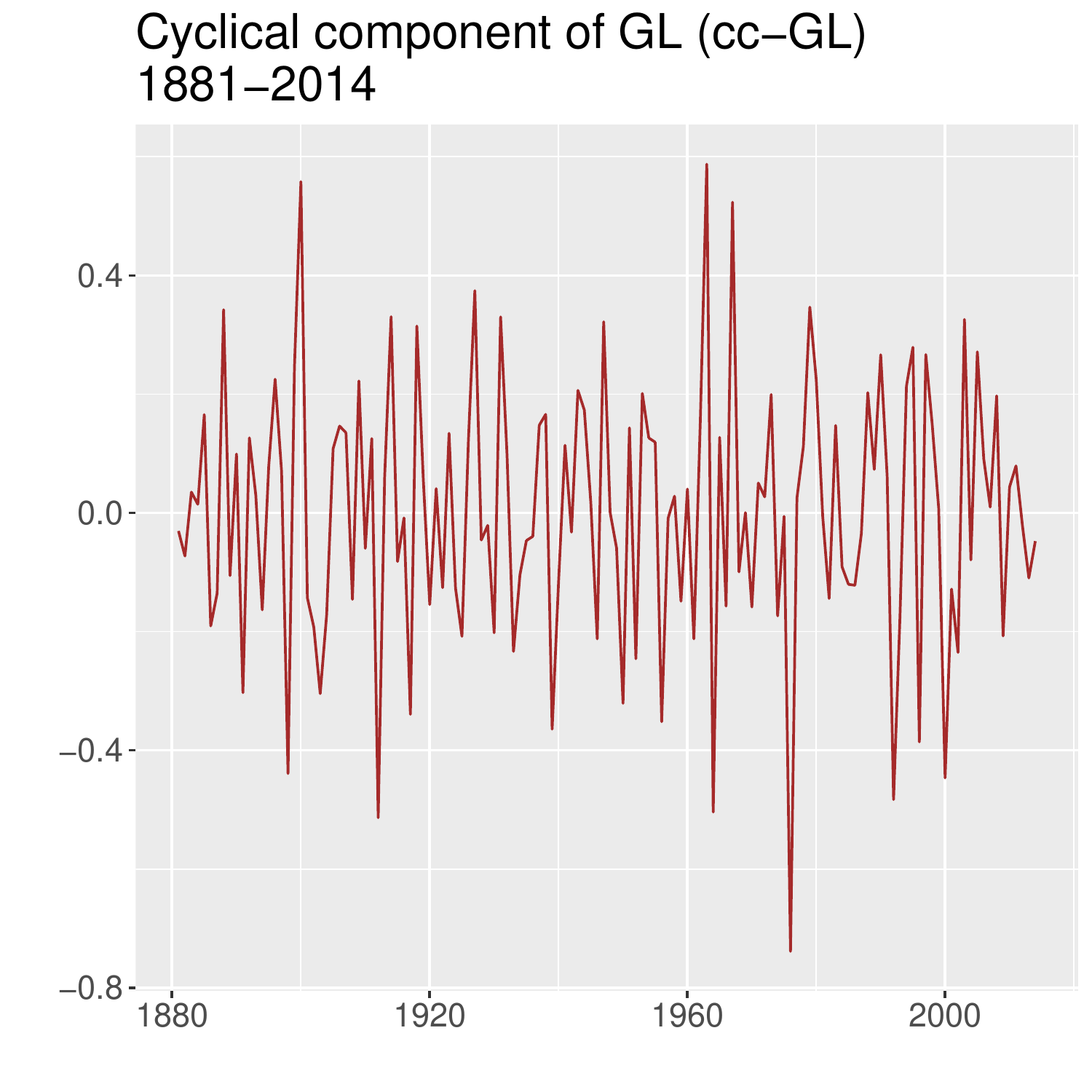}
  		 	\caption{\textit{Annual data for \\ cyclical component \\ of global land \\ temperature \\ anomalies.}}
 		\end{subfigure}
		\begin{subfigure}[t]{.24\textwidth}
   		 	\centering
   			 \includegraphics[width=\linewidth]{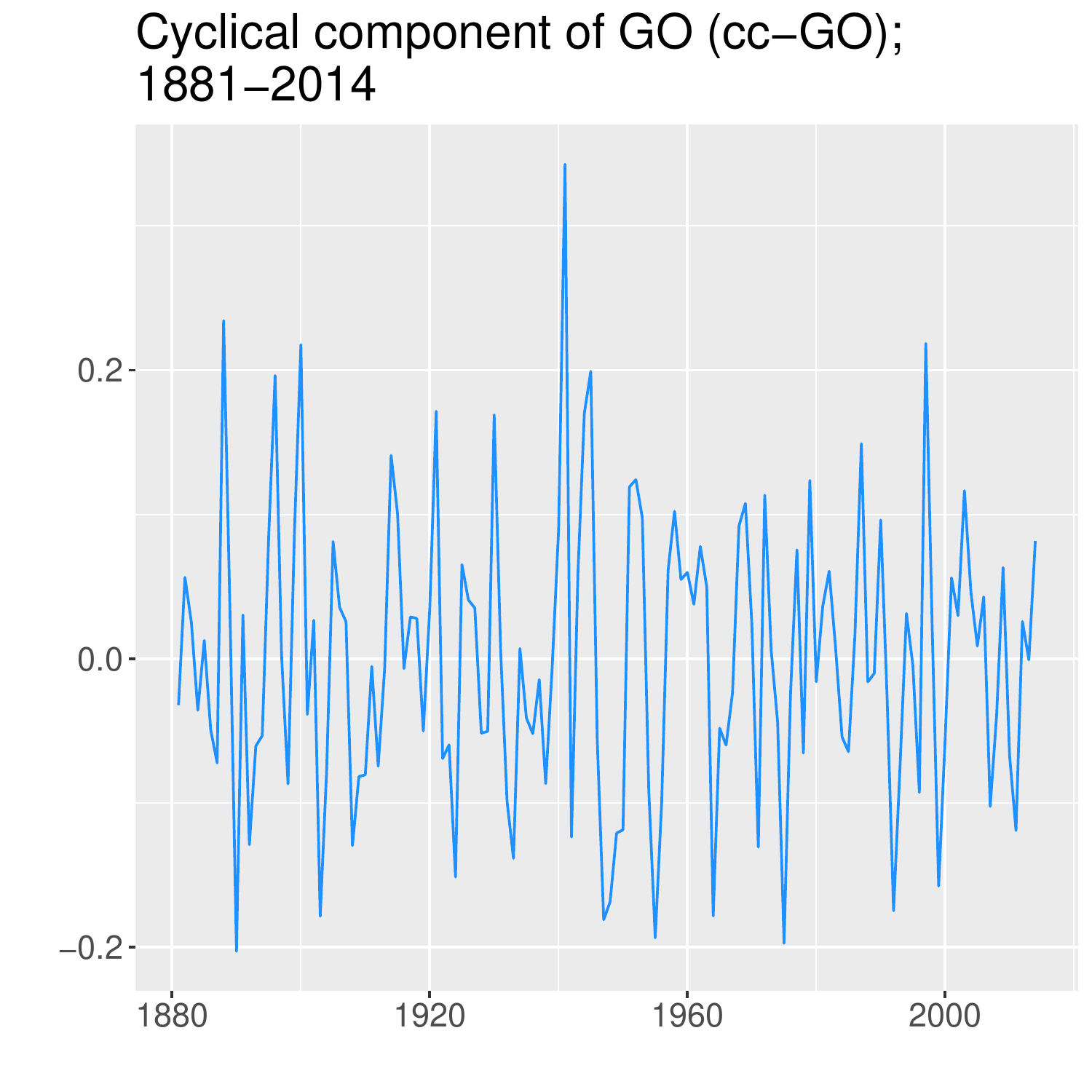}
  		 	\caption{\textit{Annual data for \\ cyclical component \\  of global ocean \\ temperature \\ anomalies.}}
 		\end{subfigure}\\
		
		\bigskip
		
		\begin{subfigure}[t]{.24\textwidth}
   		 	\centering
   			 \includegraphics[width=\linewidth]{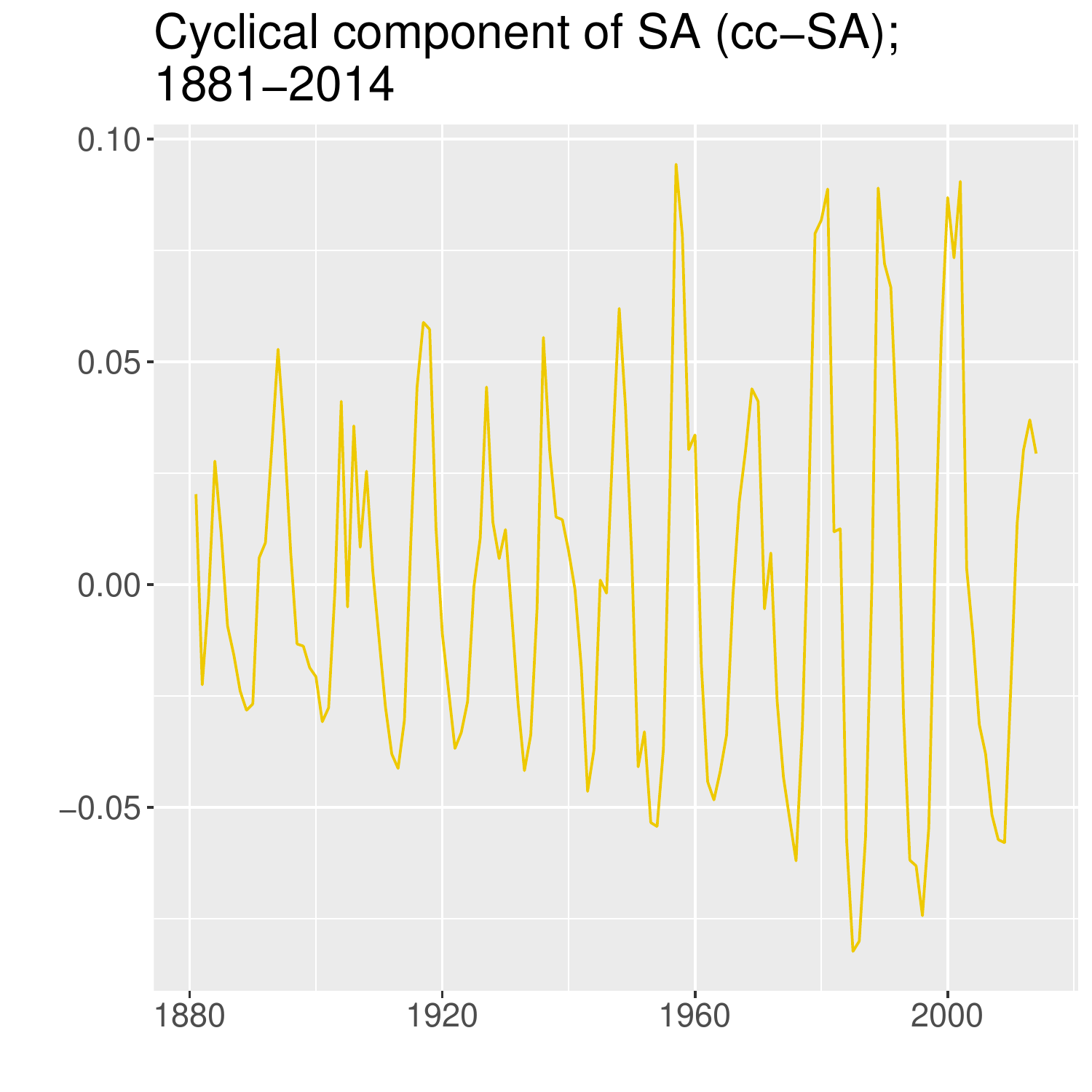}
  		 	\caption{\textit{Annual data for \\ cyclical component \\ of solar activity.}}
 		\end{subfigure}
		\begin{subfigure}[t]{.24\textwidth}
   		 	\centering
   			 \includegraphics[width=\linewidth]{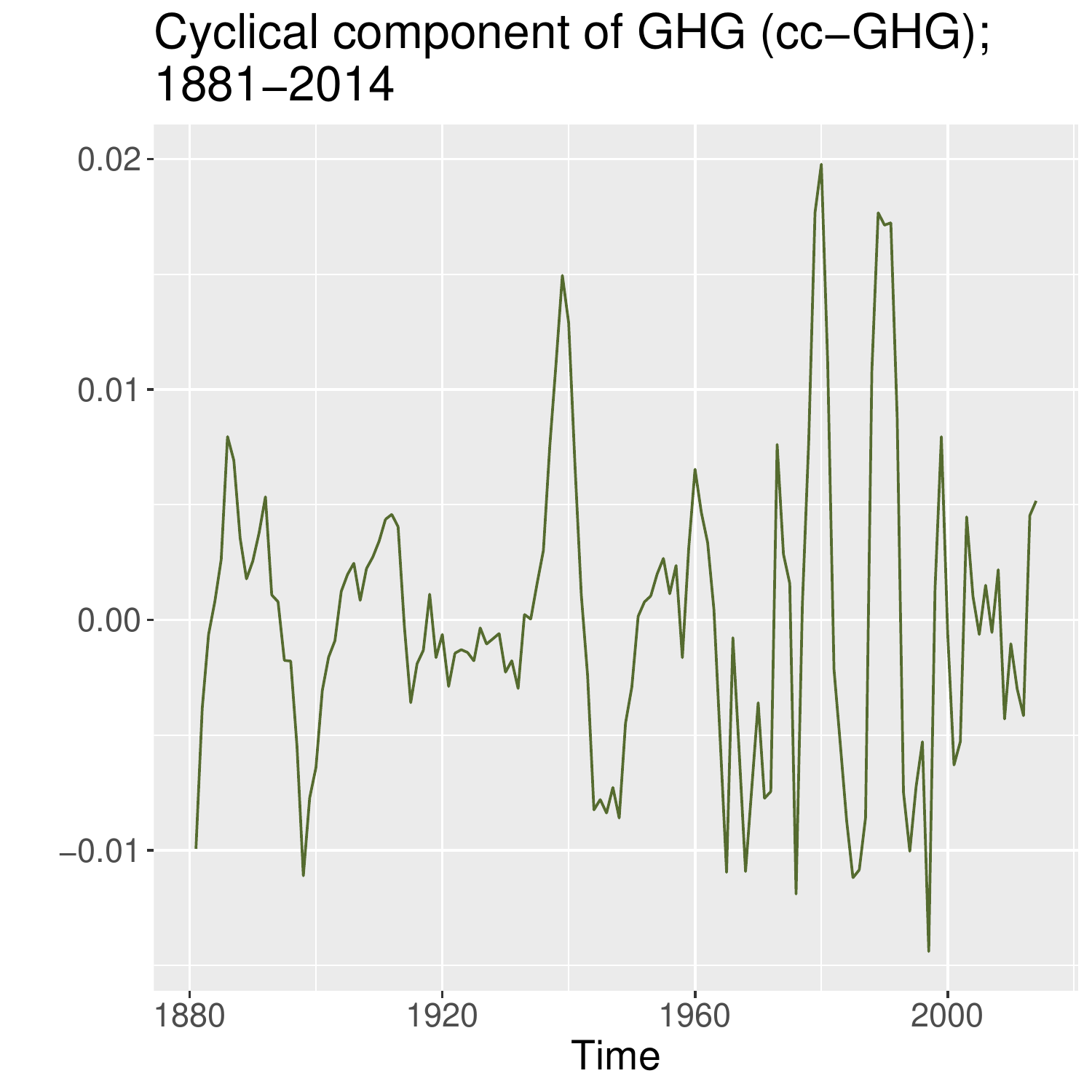}
  		 	\caption{\textit{Annual data for \\ cyclical component \\ of greenhouses gas.}}
 		\end{subfigure}
		\begin{subfigure}[t]{.24\textwidth}
   		 	\centering
   			 \includegraphics[width=\linewidth]{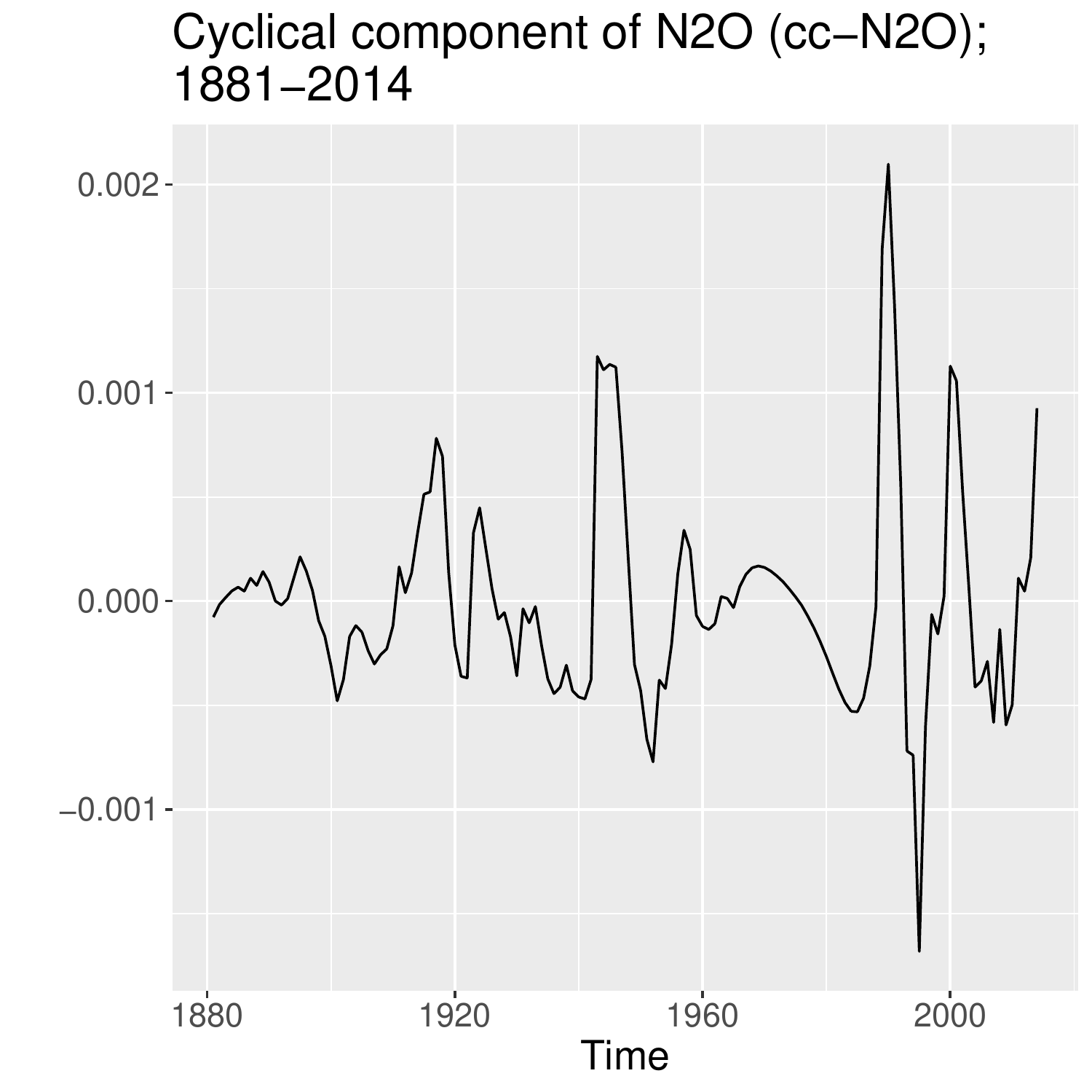}
  		 	\caption{\textit{Annual data for \\ cyclical component \\ of nitrous oxide.}}
 		\end{subfigure}
 	\caption{Cyclical components of the detrended time series.}
	
	\bigskip
		
		\begin{subfigure}[t]{.24\textwidth}
   		 	\centering
   			 \includegraphics[width=\linewidth]{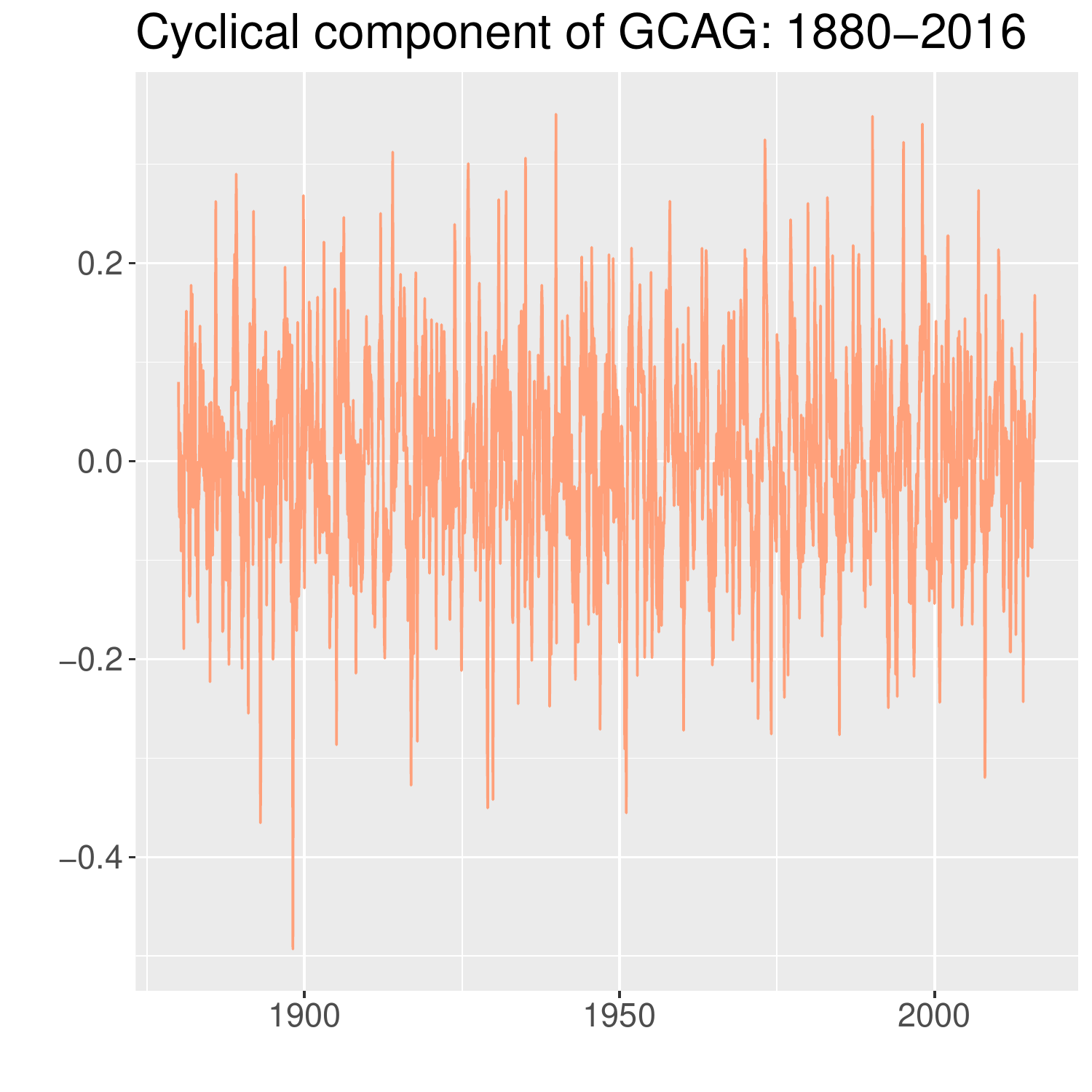}
  		 	\caption{\textit{Monthly data for \\ cyclical component of \\ global component \\ of climate at \\ a glance.}}
 		\end{subfigure}
		\begin{subfigure}[t]{.24\textwidth}
   		 	\centering
   			 \includegraphics[width=\linewidth]{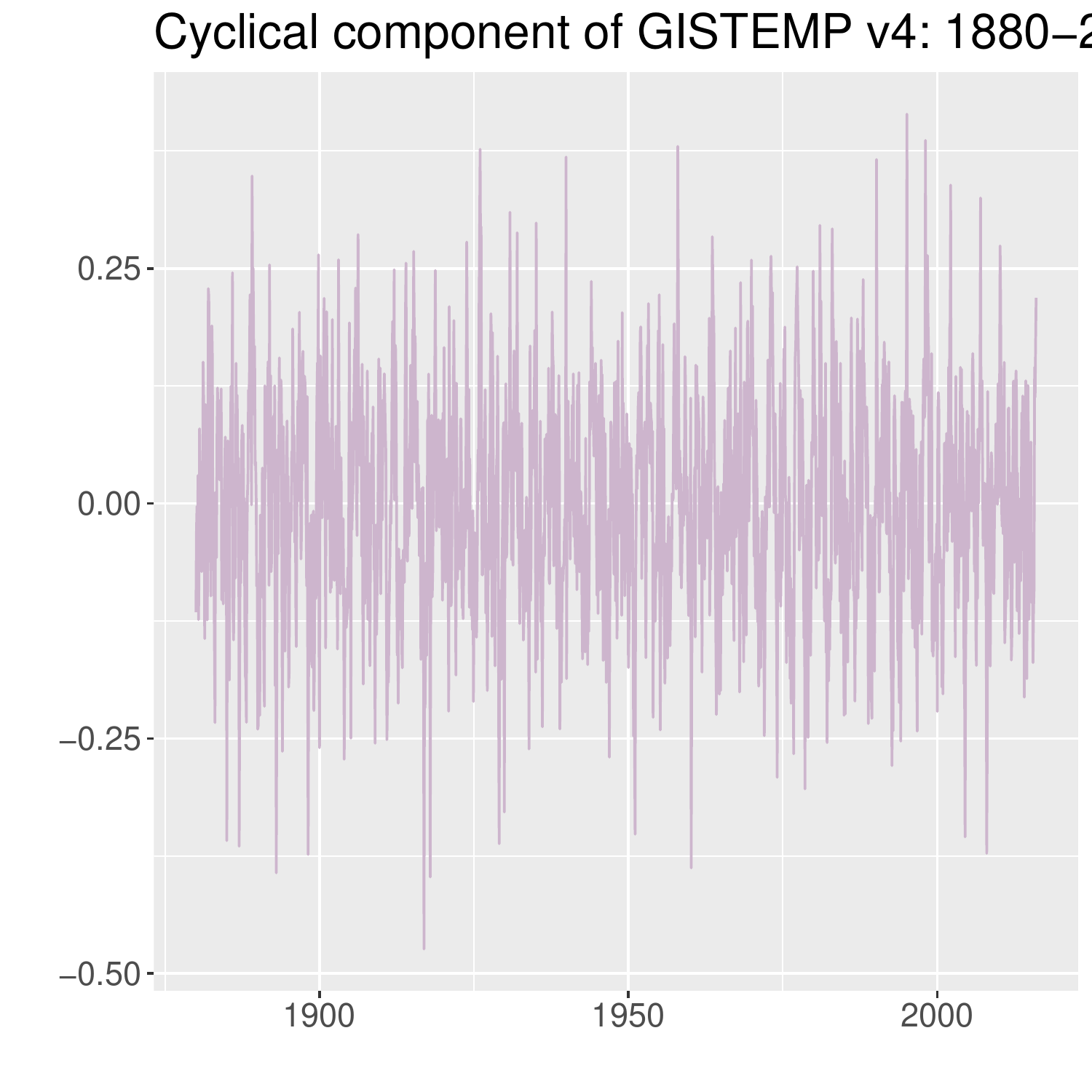}
  		 	\caption{\textit{Monthly data for \\ cyclical component \\ of global surface \\ temperature change.}}
 		\end{subfigure}
		\begin{subfigure}[t]{.24\textwidth}
   		 	\centering
   			 \includegraphics[width=\linewidth]{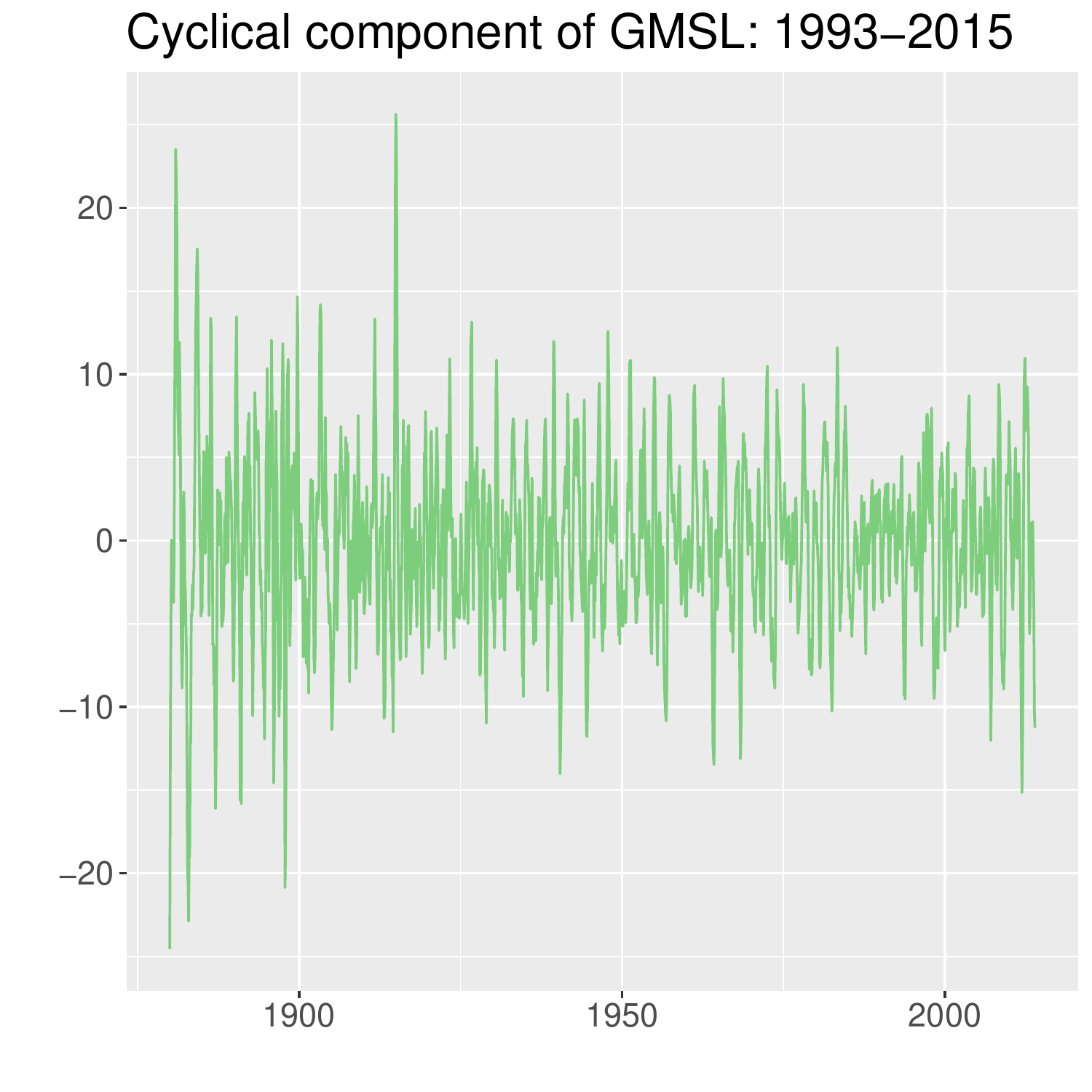}
  		 	\caption{\textit{Monthly data for \\ cyclical component of \\ global mean \\ sea level.}}
 		\end{subfigure}
	\end{center}
 	\caption{Cyclical components of the detrended time series.}
 \end{figure}
\indent 
The goal is to investigate the time-reversibility of the variables displayed in Figure 2 and the latter five in Figure 1.\\
\indent We estimate autoregressive models (see Sections 3.1 and 3.2) for each time series. We use the BIC to 
identify the number of lags ($p$). Next, we test the normality of the residuals of the nine AR($p$) models. 
Since for $cc^{GLO}$, $cc^{GL}$, $cc^{GO}$, $cc^{SA}$, and $SH$ we do not reject the null hypothesis of 
normality (significance level 0.05) of the Shapiro-Wilk test ($p$-values equal to 0.83, 0.59, 0,24, 0.08, and 
0.45, respectively) and the Jarque-Bera test ($p$-values equal to 0.64, 0.25, 0.15, 0.13, and 0.35 
respectively), we identify them as time-reversible processes. On the other hand, in $cc^{GHG}$, $cc^{N2O}$, 
$cc^{GCAG}$, $cc^{GISTEMP}$, $cc^{GMSL}$, $SOI$, $NAO$, $PDO$, and $NH$, we reject the null 
hypothesis of Gaussianity of both the Shapiro-Wilk test ($p$-values are close to zero for $cc^{GHG}$, 
$cc^{N2O}$, $cc^{GCAG}$, $cc^{GISTEMP}$, $cc^{GMSL}$, $PDO$, $NH$ and 0.0362, 0.0020 for $SOI$ 
and $NAO$, respectively) and the Jarque-Bera test ($p$-value equal to 0.0307 for $SOI$ and close to zero for 
all the other variables) at a significance level of 0.05. We can then fit MAR models to our data, identifying 
$cc^{GHG}$, $cc^{GCAG}$,and $cc^{GISTEMP}$ as MAR(2,0), $cc^{N2O}$ as MAR(4,0), $cc^{GMSL}$ 
as MAR(6,2), $PDO$ as MAR(0,4), $SOI$ as MAR(2,2), $NAO$ as MAR(1,1), and $NH$ as MAR(12,2) 
(Table 4). Since the condition $r=s$ is not met, $GHG$, $N2O$, $GCAG$, $GISTEMP$, $GMSL$, and 
$PDO$ are time-irreversible. However, TR is still a possible outcome for $SOI$ and $NAO$; hence, we 
implement the next steps of Strategies 1 and 2. Since the information criteria of the restricted MAR(2,2) 
(BIC=2705.795) is larger than the one provided by the MAR(2,2) with no restrictions (BIC=2659.974), 
Strategy 1 identifies $SOI$ as time-irreversible. The same result follows from Strategy 2: the null hypothesis 
of TR is rejected since the estimated likelihood ratio test statistic equals 52.57. Contrastingly, $NAO$ is 
identified as time-reversible from both strategies: the information criteria of the restricted MAR(1,1) is lower 
(BIC=2450.267) than the one provided by the unrestricted MAR(1,1) (BIC=2456.317), and the estimated 
likelihood ratio test statistic is equal to 0.6985. Even if this last time series rejects the null hypothesis of 
Gaussianity, it is very close to the Gaussian case since the estimated degrees of freedom ($\hat{\nu}$) 
equals 96.2. However, identifying $NAO$ as non-Gaussian does not affect our conclusions as both strategies 
identify it as time-reversible.

\begin{table}[]
\resizebox{\columnwidth}{!}{%
\begin{tabular}{@{}llllllllll@{}}
\toprule
                    & $cc^{GHG}$ & $cc^{N2O}$ & $cc^{GCAG}$ & $cc^{GISTEMP}$ & $cc^{GMSL}$ & $SOI$    & $NAO$     & $PDO$    & $NH$     \\ \midrule
$\hat{\phi}_{1}$    & 0.9620     & 0.9818     & 0.4417      & 0.4003           & 1.1233      & -0.0933  & -0.0966   & /        & 0.1657   \\
                    & (0.0841)   & (0.0722)   & (0.0225)    & (0.0239)         & (0.0231)    & (0.0327) & (0.00342) &          & (0.0300) \\
$\hat{\phi}_{2}$    & -0.3230    & -0.2413    & 0.1443      & 0.1245           & -0.1169     & -0.1315  & /         & /        & -0.0071  \\
                    & (0.0822)   & (0.0980)   & (0.0225)    & (0.0239)         & (0.0347)    & (0.0326) &           &          & (0.0306) \\
$\hat{\phi}_{3}$    & /          & -0.0016    & /           & /                & -0.6729     & /        & /         & /        & -0.0824  \\
                    &            & (0.1030)   &             &                  & (0.0336)    &          &           &          & (0.0304) \\
$\hat{\phi}_{4}$    & /          & -0.2028    & /           & /                & 0.3971      & /        & /         & /        & 0.0025   \\
                    &            & (0.0753)   &             &                  & (0.0335)    &          &           &          & (0.0303) \\
$\hat{\phi}_{5}$    & /          & /          & /           & /                & 0.0898      & /        & /         & /        & -0.0025  \\
                    &            &            &             &                  & (0.0346)    &          &           &          & (0.0303) \\
$\hat{\phi}_{6}$    & /          & /          & /           & /                & -0.1798     & /        & /         & /        & -0.0390  \\
                    &            &            &             &                  & (0.0229)    &          &           &          & (0.0303) \\
$\hat{\phi}_{7}$    & /          & /          & /           & /                & /           & /        & /         & /        & 0.0058   \\
                    &            &            &             &                  &             &          &           &          & (0.0303) \\
$\hat{\phi}_{8}$    & /          & /          & /           & /                & /           & /        & /         & /        & -0.0136  \\
                    &            &            &             &                  &             &          &           &          & (0.0303) \\
$\hat{\phi}_{9}$    & /          & /          & /           & /                & /           & /        & /         & /        & -0.0926  \\
                    &            &            &             &                  &             &          &           &          & (0.0303) \\
$\hat{\phi}_{10}$   & /          & /          & /           & /                & /           & /        & /         & /        & 0.0496   \\
                    &            &            &             &                  &             &          &           &          & (0.0304) \\
$\hat{\phi}_{11}$   & /          & /          & /           & /                & /           & /        & /         & /        & 0.1039   \\
                    &            &            &             &                  &             &          &           &          & (0.0305) \\
$\hat{\phi}_{12}$   & /          & /          & /           & /                & /           & /        & /         & /        & 0.7015   \\
                    &            &            &             &                  &             &          &           &          & (0.0300) \\ \midrule
$\hat{\varphi}_{1}$ & /          & /          & /           & /                & 0.0880      & 0.4951   & 0.2925    & 0.9183   & 0.7655   \\
                    &            &            &             &                  & (0.0225)    & (0.0313) & (0.0328)  & (0.0215) & (0.0391) \\
$\hat{\varphi}_{2}$ & /          & /          & /           & /                & 0.2709      & 0.3169   & /         & -0.1365  & -0.0512  \\
                    &            &            &             &                  & (0.0225)    & (0.0313) &           & (0.0291) & (0.0391) \\
$\hat{\varphi}_{3}$ & /          & /          & /           & /                & /           & /        & /         & 0.0063   & /        \\
                    &            &            &             &                  &             &          &           & (0.0291) &          \\
$\hat{\varphi}_{4}$ & /          & /          & /           & /                & /           & /        & /         & 0.0664   & /        \\
                    &            &            &             &                  &             &          &           & (0.0214) &          \\ \midrule
$\hat{\nu}$         & 19.8       & 13.0       & 5.3         & 9.9              & 6.8         & 8.0      & 96.2      & 9.1      & 8.2      \\ \bottomrule
\end{tabular}}
	\caption{\footnotesize{\textit{Estimated coefficients of the time series identified as non-Gaussian. The figures in parentheses are the standard errors computed by using the 
	Hessian matrix.}}}
\end{table}
In summary, our findings identify $GLO$, $GL$, $GO$, $SA$, $NAO$, and $SH$ as time-reversible and 
$GHG$, $N2O$, $GCAG$, $GISTEMP$, $GMSL$, $SOI$, $PDO$, and $NH$ as time-irreversible. The time 
irreversibility of $cc^{GHG}$ and $cc^{N2O}$ is a noticeable property of variables that account for the 
warming trend in global temperatures (\cite{IPCC14}, \cite{morana2019some}). We expect time irreversibility 
also to be present in other variables affected by greenhouse gas emissions, as, statistically, a linear 
combination of time-irreversible and time-reversible variables is also time-irreversible. This result can explain 
why $GCAG$, $GISTEMP$, $GMSL$, $SOI$, $PDO$, and $NH$ are time-irreversible. In particular, these 
results underline how global warming might have exerted feedback effects on natural oscillations, 
temperatures, and the environment in general. Among others, \cite{morana2019some} show that $GHG$ 
emissions are the key determinant of the warming trend in global temperatures. The irreversibility of 
$GCAG$ and $GISTEMP$ further corroborates these findings. Yet the evidence is inconclusive as the cyclical 
components of $GO$, $GL$, and $GLO$ are time-reversible. However, whether this latter result might be an 
artefact due to their shorter sample is plausible. \cite{morana2019some} also document Atlantic hurricanes' 
increasing natural disaster risk and destabilizing impact on the $ENSO$ cycle. Indeed, oceans warming can 
trigger a tipping point in the $ENSO$ cycle, increasing its variability and intensity and shifting its 
teleconnection eastward (\cite{cai2021changing}; see also \cite{cai2014increasing}, and 
\cite{cai2015increased}). Global warming can also profoundly affect $PDO$, shortening its lifespan and suppressing its amplitude (\cite{li2020pacific}). Moreover, the melting of land ice and warming ocean waters 
cause rising sea levels affecting coastal shorelines. High-tide flooding is increasing in magnitude and 
frequency: minor floods occur multiple times per year; major floods might occur even yearly. Even if $GHG$ 
emissions stopped, the sea level would continue to rise.  Finally, the Arctic is warming twice as faster as the 
planet on average, and finding irreversibility in $NH$ but not in $SH$, is interesting in this respect. Despite 
not being conclusive, the results might indicate that some irreversible environmental changes are ongoing.

\section{Conclusions}
This paper links the concept of an environmental tipping point to the statistical concept of time irreversibility. 
A tipping point signals an environmental change that is large, abrupt, and irreversible and generates cascading 
effects. A tipping point is a point of no return, which we associate with a temporal asymmetry in a phenomenon’s 
probabilistic structure, whereby it behaves differently according to the direction of time considered. This univocity 
along the time direction signals that the system has undergone an irreversible change. Well-known tipping points 
concern the Greenland and the West Antarctic ice sheets, the Atlantic Meridional Overturning Circulation 
($AMOC$), thawing permafrost, $ENSO$, and the Amazon rainforest. Recent $IPCC$ assessments suggest 
that tipping points might occur even between $1^{\circ}C$ and $2^{\circ}C$ warming relative to pre-industrial 
temperature averages. Therefore, they are likely to arise at current emissions levels if they have not already 
occurred.\\
\indent We then introduce two new strategies, grounded on mixed causal and noncausal models, to detect 
whether a stochastic process is time-reversible (TR). Unlike existing approaches, our methods do not impose 
strong restrictions on the model and are straightforward to implement. Moreover, similarly to 
\cite{proietti2020peaks}, they can also be applied to non-stationary processes and, therefore, useful to assess 
some key variables, such as temperature anomalies and GHG emissions, which appear to exhibit this property. 
Our simulation studies show that the strategies perform accurately and have a solid ability to detect TR.\\
\indent In the empirical analysis, we have considered fourteen climate time series, i.e., annual and monthly global 
temperature anomalies ($GLO$, $GL$, $GO$; $GCAG$, $GISTEMP$), solar activity ($SA$), natural oscillations 
($NAO$, $SOI$, $PDO$), the global mean sea level ($GMSL$), the Northern ($NH$) and Southern ($SH$) 
Hemisphere sea ice areas, global sea levels, greenhouse gas emissions ($GHG$, $N2O$). We detect time 
irreversibility in $GHG$ and $N2O$ emissions, $SOI$ and $PDO$, $GMLS$, $NH$, and the monthly temperature 
anomaly series. Yet not in the annual temperature series, $SH$, and $NAO$ (and $SA$). The time irreversibility of 
$GHG$ emissions is a noticeable property of variables that are well-known causes of global warming. It may then 
explain the time irreversibility of $GMSL$, $NH$, global temperature, and some natural oscillation indices such as 
$PDO$ and $SOI$, and therefore signal that some potentially irreversible environmental changes are ongoing. 
This evidence might not be apparent from annual temperature data due to the relatively smaller sample size 
available for annual than monthly data.\\
\indent Recent studies suggest global temperature has already warmed by $1.3^{\circ}$C and could cross the 
$1.5^{\circ}$C threshold within a decade. While not conclusive, our findings urge the implementation of correction 
policies to avoid the worst consequences of climate change and not miss the opportunity window, which might still 
be available, despite closing quickly.

\section*{Acknowledgment}
The authors would like to thank the participants in the ECˆ2 2021 (Aarhus), the CFE 2021 (London), the VI 
EMCC (Toulouse), the editor, and two anonymous referees for their valuable comments and suggestions. All remaining errors are ours.

\bibliographystyle{elsarticle-harv}\biboptions{authoryear}
\raggedright
\bibliography{IsClimateChangeTR_NewVersion.bib}

\begin{thebibliography}{41}
\expandafter\ifx\csname natexlab\endcsname\relax\def\natexlab#1{#1}\fi
\providecommand{\url}[1]{\texttt{#1}}
\providecommand{\href}[2]{#2}
\providecommand{\path}[1]{#1}
\providecommand{\DOIprefix}{doi:}
\providecommand{\ArXivprefix}{arXiv:}
\providecommand{\URLprefix}{URL: }
\providecommand{\Pubmedprefix}{pmid:}
\providecommand{\doi}[1]{\href{http://dx.doi.org/#1}{\path{#1}}}
\providecommand{\Pubmed}[1]{\href{pmid:#1}{\path{#1}}}
\providecommand{\bibinfo}[2]{#2}
\ifx\xfnm\relax \def\xfnm[#1]{\unskip,\space#1}\fi
\bibitem[{Backus and Kehoe(1992)}]{backus1992international}
\bibinfo{author}{Backus, D.K.}, \bibinfo{author}{Kehoe, P.J.},
  \bibinfo{year}{1992}.
\newblock \bibinfo{title}{International evidence on the historical properties
  of business cycles}.
\newblock \bibinfo{journal}{The American Economic Review} ,
  \bibinfo{pages}{864--888}.
\bibitem[{Belaire-Franch and Contreras(2003)}]{belaire2003tests}
\bibinfo{author}{Belaire-Franch, J.}, \bibinfo{author}{Contreras, D.},
  \bibinfo{year}{2003}.
\newblock \bibinfo{title}{Tests for time reversibility: a complementarity
  analysis}.
\newblock \bibinfo{journal}{Economics Letters} \bibinfo{volume}{81},
  \bibinfo{pages}{187--195}.
\bibitem[{Breidt and Davis(1992)}]{breidt1992time}
\bibinfo{author}{Breidt, F.J.}, \bibinfo{author}{Davis, R.A.},
  \bibinfo{year}{1992}.
\newblock \bibinfo{title}{Time-reversibility, identifiability and independence
  of innovations for stationary time series}.
\newblock \bibinfo{journal}{Journal of Time Series Analysis}
  \bibinfo{volume}{13}, \bibinfo{pages}{377--390}.
\bibitem[{Breidt et~al.(1991)Breidt, Davis, Lh and
  Rosenblatt}]{breid1991maximum}
\bibinfo{author}{Breidt, F.J.}, \bibinfo{author}{Davis, R.A.},
  \bibinfo{author}{Lh, K.S.}, \bibinfo{author}{Rosenblatt, M.},
  \bibinfo{year}{1991}.
\newblock \bibinfo{title}{Maximum likelihood estimation for noncausal
  autoregressive processes}.
\newblock \bibinfo{journal}{Journal of Multivariate Analysis}
  \bibinfo{volume}{36}, \bibinfo{pages}{175--198}.
\bibitem[{Caesar et~al.(2021)Caesar, McCarthy, Thornalley, Cahill and
  Rahmstorf}]{caesar2021current}
\bibinfo{author}{Caesar, L.}, \bibinfo{author}{McCarthy, G.},
  \bibinfo{author}{Thornalley, D.}, \bibinfo{author}{Cahill, N.},
  \bibinfo{author}{Rahmstorf, S.}, \bibinfo{year}{2021}.
\newblock \bibinfo{title}{Current {Atlantic} {Meridional} {Overturning}
  {Circulation} weakest in last millennium}.
\newblock \bibinfo{journal}{Nature Geoscience} \bibinfo{volume}{14},
  \bibinfo{pages}{118--120}.
\bibitem[{Cai et~al.(2014)Cai, Borlace, Lengaigne, Van~Rensch, Collins, Vecchi,
  Timmermann, Santoso, McPhaden, Wu et~al.}]{cai2014increasing}
\bibinfo{author}{Cai, W.}, \bibinfo{author}{Borlace, S.},
  \bibinfo{author}{Lengaigne, M.}, \bibinfo{author}{Van~Rensch, P.},
  \bibinfo{author}{Collins, M.}, \bibinfo{author}{Vecchi, G.},
  \bibinfo{author}{Timmermann, A.}, \bibinfo{author}{Santoso, A.},
  \bibinfo{author}{McPhaden, M.J.}, \bibinfo{author}{Wu, L.}, et~al.,
  \bibinfo{year}{2014}.
\newblock \bibinfo{title}{Increasing frequency of extreme el ni{\~n}o events
  due to greenhouse warming}.
\newblock \bibinfo{journal}{Nature climate change} \bibinfo{volume}{4},
  \bibinfo{pages}{111--116}.
\bibitem[{Cai et~al.(2021)Cai, Santoso, Collins, Dewitte, Karamperidou, Kug,
  Lengaigne, McPhaden, Stuecker, Taschetto et~al.}]{cai2021changing}
\bibinfo{author}{Cai, W.}, \bibinfo{author}{Santoso, A.},
  \bibinfo{author}{Collins, M.}, \bibinfo{author}{Dewitte, B.},
  \bibinfo{author}{Karamperidou, C.}, \bibinfo{author}{Kug, J.S.},
  \bibinfo{author}{Lengaigne, M.}, \bibinfo{author}{McPhaden, M.J.},
  \bibinfo{author}{Stuecker, M.F.}, \bibinfo{author}{Taschetto, A.S.}, et~al.,
  \bibinfo{year}{2021}.
\newblock \bibinfo{title}{Changing {El} {Ni{\~n}o}--{Southern} {Oscillation} in
  a warming climate}.
\newblock \bibinfo{journal}{Nature Reviews Earth \& Environment}
  \bibinfo{volume}{2}, \bibinfo{pages}{628--644}.
\bibitem[{Cai et~al.(2015)Cai, Wang, Santoso, McPhaden, Wu, Jin, Timmermann,
  Collins, Vecchi, Lengaigne et~al.}]{cai2015increased}
\bibinfo{author}{Cai, W.}, \bibinfo{author}{Wang, G.},
  \bibinfo{author}{Santoso, A.}, \bibinfo{author}{McPhaden, M.J.},
  \bibinfo{author}{Wu, L.}, \bibinfo{author}{Jin, F.F.},
  \bibinfo{author}{Timmermann, A.}, \bibinfo{author}{Collins, M.},
  \bibinfo{author}{Vecchi, G.}, \bibinfo{author}{Lengaigne, M.}, et~al.,
  \bibinfo{year}{2015}.
\newblock \bibinfo{title}{Increased frequency of extreme la ni{\~n}a events
  under greenhouse warming}.
\newblock \bibinfo{journal}{Nature Climate Change} \bibinfo{volume}{5},
  \bibinfo{pages}{132--137}.
\bibitem[{Chen et~al.(2000)Chen, Chou and Kuan}]{chen2000testing}
\bibinfo{author}{Chen, Y.T.}, \bibinfo{author}{Chou, R.Y.},
  \bibinfo{author}{Kuan, C.M.}, \bibinfo{year}{2000}.
\newblock \bibinfo{title}{Testing time reversibility without moment
  restrictions}.
\newblock \bibinfo{journal}{Journal of Econometrics} \bibinfo{volume}{95},
  \bibinfo{pages}{199--218}.
\bibitem[{De~Jong and Sakarya(2016)}]{de2016econometrics}
\bibinfo{author}{De~Jong, R.M.}, \bibinfo{author}{Sakarya, N.},
  \bibinfo{year}{2016}.
\newblock \bibinfo{title}{The econometrics of the hodrick-prescott filter}.
\newblock \bibinfo{journal}{Review of Economics and Statistics}
  \bibinfo{volume}{98}, \bibinfo{pages}{310--317}.
\bibitem[{DeConto et~al.(2021)DeConto, Pollard, Alley
  et~al.}]{deconto2021paris}
\bibinfo{author}{DeConto, R.M.}, \bibinfo{author}{Pollard, D.},
  \bibinfo{author}{Alley, R.B.}, et~al., \bibinfo{year}{2021}.
\newblock \bibinfo{title}{The {Paris} {Climate} {Agreement} and future
  sea-level rise from {Antarctica}}.
\newblock \bibinfo{journal}{Nature} \bibinfo{volume}{593},
  \bibinfo{pages}{83--89}.
\bibitem[{Fern{\'a}ndez and Steel(1998)}]{fernandez1998bayesian}
\bibinfo{author}{Fern{\'a}ndez, C.}, \bibinfo{author}{Steel, M.F.},
  \bibinfo{year}{1998}.
\newblock \bibinfo{title}{On bayesian modeling of fat tails and skewness}.
\newblock \bibinfo{journal}{Journal of the american statistical association}
  \bibinfo{volume}{93}, \bibinfo{pages}{359--371}.
\bibitem[{Fries(2021)}]{fries2021conditional}
\bibinfo{author}{Fries, S.}, \bibinfo{year}{2021}.
\newblock \bibinfo{title}{Conditional moments of noncausal alpha-stable
  processes and the prediction of bubble crash odds}.
\newblock \bibinfo{journal}{Journal of Business \& Economic Statistics} ,
  \bibinfo{pages}{1--37}.
\bibitem[{Fries and Zakoian(2019)}]{fries2019mixed}
\bibinfo{author}{Fries, S.}, \bibinfo{author}{Zakoian, J.M.},
  \bibinfo{year}{2019}.
\newblock \bibinfo{title}{Mixed causal-noncausal ar processes and the modelling
  of explosive bubbles}.
\newblock \bibinfo{journal}{Econometric Theory} \bibinfo{volume}{35},
  \bibinfo{pages}{1234--1270}.
\bibitem[{Giancaterini and Hecq(2022)}]{GIANCATERINI2022}
\bibinfo{author}{Giancaterini, F.}, \bibinfo{author}{Hecq, A.},
  \bibinfo{year}{2022}.
\newblock \bibinfo{title}{Inference in mixed causal and noncausal models with
  generalized student’s t-distributions}.
\newblock \bibinfo{journal}{Econometrics and Statistics,
  https://doi.org/10.1016/j.ecosta.2021.11.007} .
\bibitem[{Gourieroux and Jasiak(2016)}]{gourieroux2016filtering}
\bibinfo{author}{Gourieroux, C.}, \bibinfo{author}{Jasiak, J.},
  \bibinfo{year}{2016}.
\newblock \bibinfo{title}{Filtering, prediction and simulation methods for
  noncausal processes}.
\newblock \bibinfo{journal}{Journal of Time Series Analysis}
  \bibinfo{volume}{37}, \bibinfo{pages}{405--430}.
\bibitem[{Gourieroux and Jasiak(2022)}]{gourieroux2022nonlinear}
\bibinfo{author}{Gourieroux, C.}, \bibinfo{author}{Jasiak, J.},
  \bibinfo{year}{2022}.
\newblock \bibinfo{title}{Nonlinear forecasts and impulse responses for
  causal-noncausal (s) var models}.
\newblock \bibinfo{journal}{Manuscript, University of Toronto} .
\bibitem[{Gouri{\'e}roux et~al.(2013)Gouri{\'e}roux, Zakoian
  et~al.}]{gourieroux2013explosive}
\bibinfo{author}{Gouri{\'e}roux, C.}, \bibinfo{author}{Zakoian, J.M.}, et~al.,
  \bibinfo{year}{2013}.
\newblock \bibinfo{title}{Explosive bubble modelling by noncausal process}.
\newblock \bibinfo{publisher}{CREST}.
\bibitem[{Hallin et~al.(1988)Hallin, Lefevre and Puri}]{hallin1988time}
\bibinfo{author}{Hallin, M.}, \bibinfo{author}{Lefevre, C.},
  \bibinfo{author}{Puri, M.L.}, \bibinfo{year}{1988}.
\newblock \bibinfo{title}{On time-reversibility and the uniqueness of moving
  average representations for non-gaussian stationary time series}.
\newblock \bibinfo{journal}{Biometrika} \bibinfo{volume}{75},
  \bibinfo{pages}{170--171}.
\bibitem[{Hansen et~al.(2017)Hansen, Sato, Kharecha, Von~Schuckmann, Beerling,
  Cao, Marcott, Masson-Delmotte, Prather, Rohling et~al.}]{hansen2017young}
\bibinfo{author}{Hansen, J.}, \bibinfo{author}{Sato, M.},
  \bibinfo{author}{Kharecha, P.}, \bibinfo{author}{Von~Schuckmann, K.},
  \bibinfo{author}{Beerling, D.J.}, \bibinfo{author}{Cao, J.},
  \bibinfo{author}{Marcott, S.}, \bibinfo{author}{Masson-Delmotte, V.},
  \bibinfo{author}{Prather, M.J.}, \bibinfo{author}{Rohling, E.J.}, et~al.,
  \bibinfo{year}{2017}.
\newblock \bibinfo{title}{Young people's burden: requirement of negative co 2
  emissions}.
\newblock \bibinfo{journal}{Earth System Dynamics} \bibinfo{volume}{8},
  \bibinfo{pages}{577--616}.
\bibitem[{Hecq et~al.(2016)Hecq, Lieb and Telg}]{hecq2016identification}
\bibinfo{author}{Hecq, A.}, \bibinfo{author}{Lieb, L.}, \bibinfo{author}{Telg,
  S.}, \bibinfo{year}{2016}.
\newblock \bibinfo{title}{Identification of mixed causal-noncausal models in
  finite samples}.
\newblock \bibinfo{journal}{Annals of Economics and Statistics/Annales
  d'{\'E}conomie et de Statistique} , \bibinfo{pages}{307--331}.
\bibitem[{Hecq and Voisin(2021)}]{hecq21predicting}
\bibinfo{author}{Hecq, A.}, \bibinfo{author}{Voisin, E.}, \bibinfo{year}{2021}.
\newblock \bibinfo{title}{Predicting bubble bursts in oil prices using mixed
  causal-noncausal models}.
\newblock \bibinfo{journal}{Forthcoming in Advances in Econometrics in honor of
  Joon Y. Park} .
\bibitem[{Hencic and Gouri{\'e}roux(2015)}]{hencic2015noncausal}
\bibinfo{author}{Hencic, A.}, \bibinfo{author}{Gouri{\'e}roux, C.},
  \bibinfo{year}{2015}.
\newblock \bibinfo{title}{Noncausal autoregressive model in application to
  bitcoin/usd exchange rates}, in: \bibinfo{booktitle}{Econometrics of risk}.
  \bibinfo{publisher}{Springer}, pp. \bibinfo{pages}{17--40}.
\bibitem[{Hinich and Rothman(1998)}]{hinich1998frequency}
\bibinfo{author}{Hinich, M.J.}, \bibinfo{author}{Rothman, P.},
  \bibinfo{year}{1998}.
\newblock \bibinfo{title}{Frequency-domain test of time reversibility}.
\newblock \bibinfo{journal}{Macroeconomic Dynamics} \bibinfo{volume}{2},
  \bibinfo{pages}{72--88}.
\bibitem[{Holster(2003)}]{holster2003criterion}
\bibinfo{author}{Holster, A.}, \bibinfo{year}{2003}.
\newblock \bibinfo{title}{The criterion for time symmetry of probabilistic
  theories and the reversibility of quantum mechanics}.
\newblock \bibinfo{journal}{New Journal of Physics} \bibinfo{volume}{5},
  \bibinfo{pages}{130}.
\bibitem[{IPCC(2014)}]{IPCC14}
\bibinfo{author}{IPCC}, \bibinfo{year}{2014}.
\newblock \bibinfo{title}{International panel on climate change fifth
  assessment report, available at https://www.ipcc.ch/report/ar5/syr/} .
\bibitem[{IPCC(2022)}]{IPCC22}
\bibinfo{author}{IPCC}, \bibinfo{year}{2022}.
\newblock \bibinfo{title}{International panel on climate change sixth
  assessment report, available at
  https://www.ipcc.ch/report/sixth-assessment-report-cycle/} .
\bibitem[{Lanne and Saikkonen(2011)}]{lanne2011noncausal}
\bibinfo{author}{Lanne, M.}, \bibinfo{author}{Saikkonen, P.},
  \bibinfo{year}{2011}.
\newblock \bibinfo{title}{Noncausal autoregressions for economic time series}.
\newblock \bibinfo{journal}{Journal of Time Series Econometrics}
  \bibinfo{volume}{3}.
\bibitem[{Levesque and Verlet(1993)}]{levesque1993molecular}
\bibinfo{author}{Levesque, D.}, \bibinfo{author}{Verlet, L.},
  \bibinfo{year}{1993}.
\newblock \bibinfo{title}{Molecular dynamics and time reversibility}.
\newblock \bibinfo{journal}{Journal of Statistical Physics}
  \bibinfo{volume}{72}, \bibinfo{pages}{519--537}.
\bibitem[{Li et~al.(2020)Li, Wu, Yang, Geng, Cai, Gan, Chen, Jing, Wang and
  Ma}]{li2020pacific}
\bibinfo{author}{Li, S.}, \bibinfo{author}{Wu, L.}, \bibinfo{author}{Yang, Y.},
  \bibinfo{author}{Geng, T.}, \bibinfo{author}{Cai, W.}, \bibinfo{author}{Gan,
  B.}, \bibinfo{author}{Chen, Z.}, \bibinfo{author}{Jing, Z.},
  \bibinfo{author}{Wang, G.}, \bibinfo{author}{Ma, X.}, \bibinfo{year}{2020}.
\newblock \bibinfo{title}{The pacific decadal oscillation less predictable
  under greenhouse warming}.
\newblock \bibinfo{journal}{Nature Climate Change} \bibinfo{volume}{10},
  \bibinfo{pages}{30--34}.
\bibitem[{Lovejoy and Nobre(2018)}]{lovejoy2018amazon}
\bibinfo{author}{Lovejoy, T.E.}, \bibinfo{author}{Nobre, C.},
  \bibinfo{year}{2018}.
\newblock \bibinfo{title}{Amazon {Tipping} {Point}}.
\newblock \bibinfo{journal}{Science Advances} \bibinfo{volume}{4},
  \bibinfo{pages}{eaat2340}.
\bibitem[{Meinshausen et~al.(2017)Meinshausen, Vogel, Nauels, Lorbacher,
  Meinshausen, Etheridge, Fraser, Montzka, Rayner, Trudinger
  et~al.}]{meinshausen2017historical}
\bibinfo{author}{Meinshausen, M.}, \bibinfo{author}{Vogel, E.},
  \bibinfo{author}{Nauels, A.}, \bibinfo{author}{Lorbacher, K.},
  \bibinfo{author}{Meinshausen, N.}, \bibinfo{author}{Etheridge, D.M.},
  \bibinfo{author}{Fraser, P.J.}, \bibinfo{author}{Montzka, S.A.},
  \bibinfo{author}{Rayner, P.J.}, \bibinfo{author}{Trudinger, C.M.}, et~al.,
  \bibinfo{year}{2017}.
\newblock \bibinfo{title}{Historical greenhouse gas concentrations for climate
  modelling (cmip6)}.
\newblock \bibinfo{journal}{Geoscientific Model Development}
  \bibinfo{volume}{10}, \bibinfo{pages}{2057--2116}.
\bibitem[{Morana and Sbrana(2019)}]{morana2019some}
\bibinfo{author}{Morana, C.}, \bibinfo{author}{Sbrana, G.},
  \bibinfo{year}{2019}.
\newblock \bibinfo{title}{Some financial implications of global warming: An
  empirical assessment}.
\newblock \bibinfo{journal}{Economic Modelling} \bibinfo{volume}{81},
  \bibinfo{pages}{274--294}.
\bibitem[{Proietti(2020)}]{proietti2020peaks}
\bibinfo{author}{Proietti, T.}, \bibinfo{year}{2020}.
\newblock \bibinfo{title}{Peaks, gaps, and time reversibility of economic time
  series}.
\newblock \bibinfo{journal}{WP Rome} .
\bibitem[{Ramsey and Rothman(1996)}]{ramsey1996time}
\bibinfo{author}{Ramsey, J.B.}, \bibinfo{author}{Rothman, P.},
  \bibinfo{year}{1996}.
\newblock \bibinfo{title}{Time irreversibility and business cycle asymmetry}.
\newblock \bibinfo{journal}{Journal of Money, Credit and Banking}
  \bibinfo{volume}{28}, \bibinfo{pages}{1--21}.
\bibitem[{Ravn and Uhlig(2002)}]{ravn2002adjusting}
\bibinfo{author}{Ravn, M.O.}, \bibinfo{author}{Uhlig, H.},
  \bibinfo{year}{2002}.
\newblock \bibinfo{title}{On adjusting the hodrick-prescott filter for the
  frequency of observations}.
\newblock \bibinfo{journal}{Review of economics and statistics}
  \bibinfo{volume}{84}, \bibinfo{pages}{371--376}.
\bibitem[{Schellnhuber(2008)}]{schellnhuber2008global}
\bibinfo{author}{Schellnhuber, H.J.}, \bibinfo{year}{2008}.
\newblock \bibinfo{title}{Global warming: Stop worrying, start panicking?}
\newblock \bibinfo{journal}{Proceedings of the National Academy of Sciences}
  \bibinfo{volume}{105}, \bibinfo{pages}{14239--14240}.
\bibitem[{Solomon et~al.(2009)Solomon, Plattner, Knutti and
  Friedlingstein}]{solomon2009irreversible}
\bibinfo{author}{Solomon, S.}, \bibinfo{author}{Plattner, G.K.},
  \bibinfo{author}{Knutti, R.}, \bibinfo{author}{Friedlingstein, P.},
  \bibinfo{year}{2009}.
\newblock \bibinfo{title}{Irreversible climate change due to carbon dioxide
  emissions}.
\newblock \bibinfo{journal}{Proceedings of the national academy of sciences}
  \bibinfo{volume}{106}, \bibinfo{pages}{1704--1709}.
\bibitem[{Wald(1980)}]{wald1980quantum}
\bibinfo{author}{Wald, R.M.}, \bibinfo{year}{1980}.
\newblock \bibinfo{title}{Quantum gravity and time reversibility}.
\newblock \bibinfo{journal}{Physical Review D} \bibinfo{volume}{21},
  \bibinfo{pages}{2742}.
\bibitem[{Weiss(1975)}]{weiss1975time}
\bibinfo{author}{Weiss, G.}, \bibinfo{year}{1975}.
\newblock \bibinfo{title}{Time-reversibility of linear stochastic processes}.
\newblock \bibinfo{journal}{Journal of Applied Probability}
  \bibinfo{volume}{12}, \bibinfo{pages}{831--836}.
\bibitem[{Wunderling et~al.(2021)Wunderling, Donges, Kurths and
  Winkelmann}]{wunderling2021interacting}
\bibinfo{author}{Wunderling, N.}, \bibinfo{author}{Donges, J.F.},
  \bibinfo{author}{Kurths, J.}, \bibinfo{author}{Winkelmann, R.},
  \bibinfo{year}{2021}.
\newblock \bibinfo{title}{Interacting tipping elements increase risk of climate
  domino effects under global warming}.
\newblock \bibinfo{journal}{Earth System Dynamics} \bibinfo{volume}{12},
  \bibinfo{pages}{601--619}.

\end{thebibliography}
\end{document}